\documentclass[prl,a4paper,11pt]{article}
\pdfoutput=1

\usepackage{bm}        
\usepackage{amsmath}						
\usepackage{amssymb}						
\usepackage{amsthm}						
\usepackage{array}							
\usepackage[english]{babel}					
\usepackage{cancel}						
\usepackage{eufrak}							
\usepackage{fullpage}						
\usepackage{graphicx}						
\usepackage{multirow}						
\usepackage{nccmath}						
\usepackage{nicefrac}						
\usepackage[section]{placeins}                         
\usepackage{booktabs}						
\usepackage[format=hang]{caption}				
\usepackage[hang,flushmargin]{footmisc}		
\usepackage{hyperref}						
\usepackage{mathtools}						

\numberwithin{equation}{section}				
\newcommand{\balpha}{\bar{\alpha}}			
\newcommand{\bbeta}{\bar{\beta}}
\newcommand{\bgamma}{\bar{\gamma}}
\newcommand{\bdelta}{\bar{\delta}}
\newcommand{\bepsilon}{\bar{\epsilon}}
\newcommand{\bmu}{\bar{\mu}}
\newcommand{\bpi}{\bar{\pi}}
\newcommand{\brho}{\bar{\rho}}
\newcommand{\btau}{\bar{\tau}}

\newcommand{\teta}{\tilde{\eta}}				
\newcommand{\tLambda}{\widetilde{\Lambda}}

\newcommand{\NPl}{\bm{\ell}}					
\newcommand{\NPn}{\bm{n}}
\newcommand{\NPm}{\bm{m}}
\newcommand{\NPbm}{\bm{\bar{m}}}

\newcommand{\Naturals}{\mathbb{N}}			
\newcommand{\Integers}{\mathbb{Z}}			
\newcommand{\Reals}{\mathbb{R}}				
\newcommand{\Imaginaries}{\mathbb{I}}			

\begin{document}

\thispagestyle{empty}

\vspace*{2cm}

\begin{center}
{\bf \LARGE Kerr-AdS and its Near-horizon Geometry:\\
Perturbations and the Kerr/CFT Correspondence}\\
\vspace*{2.5cm}

{\bf \'Oscar J. C.~Dias$^\star$, } {\bf Jorge E.~Santos$^\ddagger$, } {\bf Maren Stein$^\dagger$ }

\vspace*{1cm}

 {\it $\star$ Institut de Physique Th\'eorique, CEA Saclay,\\
 CNRS URA 2306, F-91191 Gif-sur-Yvette, France}\\ \vspace{0.3cm}
 {\it $\ddagger$ Department of Physics, UCSB, Santa Barbara, CA 93106, USA}\\ \vspace{0.3cm}
 {\it $\dagger$ DAMTP, Centre for Mathematical Sciences, University of Cambridge,\\
 Wilberforce Road, Cambridge CB3 0WA, United Kingdom}

\vspace*{0.5cm} {\tt oscar.dias@cea.fr, jss55@physics.ucsb.edu, mcs60@cam.ac.uk}

\end{center}

\vspace*{1cm}

\begin{abstract}
We investigate linear perturbations of spin-$s$ fields in the Kerr-AdS black hole and in its near-horizon geometry (NHEK-AdS), using the Teukolsky master equation and the Hertz potential. In the NHEK-AdS geometry we solve the associated angular equation numerically and the radial equation exactly.  Having these explicit solutions at hand, we search for linear mode instabilities. We do not find any (non-)axisymmetric instabilities with outgoing boundary conditions. This is in agreement with a recent conjecture relating the linearized stability properties of the full geometry with those of its near-horizon geometry. Moreover, we find that the asymptotic behaviour of the metric perturbations in NHEK-AdS violates the fall-off conditions imposed in the formulation of the Kerr/CFT correspondence (the only exception being the axisymmetric sector of perturbations).
\end{abstract}
\noindent

\newpage
\thispagestyle{empty}
\tableofcontents

\setcounter{page}{0} \setcounter{footnote}{0}

\section{Introduction and summary\label{IntroSum}}

The Kerr black hole is the unique black hole solution in the phase diagram of stationary solutions of $d=4$ asymptotically flat Einstein gravity and, ultimately,  it describes an isolated astrophysical black hole. Therefore it is reassuring that Whiting \cite{Whiting:1988vc}, using the results of Press and Teukolsky \cite{Press:1973zz}, found that the Kerr solution is linearly stable in a mode by mode analysis of linearized non-algebraically-special gravitational perturbations.  Technically, this analysis was possible due to the Newman-Penrose formalism whereby all the gravitational perturbation information is encoded in two decoupled complex Weyl scalars. These are gauge invariant quantities with the same number of degrees of freedom as the metric perturbation. Moreover, Teukolsky \cite{Teukolsky:1973ha} proved that there is a single decoupled master equation governing the perturbations of these Weyl scalars. In a mode by mode analysis, this master equation further separates into a radial and angular equation which makes the analysis tractable. An interesting property of the Kerr black hole is that it has an extreme configuration where the temperature vanishes but its entropy remains finite. Bardeen and Horowitz \cite{Bardeen:1999px} described how one can take a near-horizon limit of this extreme Kerr geometry to get a spacetime similar to $AdS_2 \times S^2$ that is called the near-horizon extreme Kerr geometry (NHEK). The naive intuition suggests that a necessary but not sufficient condition for the stability of the (near-)extreme Kerr solution is then that NHEK itself should be stable subject to appropriate boundary conditions. Refs. \cite{Amsel:2009ev,Dias:2009ex} found that NHEK is linearly stable in a mode by mode analysis. 
We emphasize the fact that the above condition is necessary but not sufficient. That is, we  can have a linear instability of the full extreme Kerr geometry $-$ see \cite{Marolf:2010nd,Aretakis:2012ei,LuciettiReall2012} $-$ that is however not captured by a linear instability analysis of NHEK \cite{Amsel:2009ev,Dias:2009ex}. 

In an asymptotically anti-de Sitter (AdS) background, the Kerr-AdS black hole is the only  stationary black hole of $d=4$ Einstein-AdS theory whose solution is exactly known \cite{Carter:1968ks}.\footnote{There is perturbative evidence that it might not be the only stationary black hole of the theory. Indeed, Ref. \cite{Dias:2011ss} constructed perturbatively a rotating black hole with a single Killing vector field by placing a Kerr-AdS black at the core of a geon.}       
These black holes are linearly unstable (at least) to the (non-axisymmetric) gravitational superradiant instability if their angular velocity is larger than $1$ in AdS units \cite{Hawking:1999dp,Cardoso:2006wa}. Again, this conclusion can be achieved solving the Teukolsky master equation in the Kerr-AdS black hole. Indeed, this equation can be derived as long as the background is Petrov type D, with Kerr(-AdS) and their near-horizon geometries being in this category. The extreme Kerr-AdS black hole also has a near-horizon geometry $-$ the NHEK-AdS $-$ explicitly derived by L\"u, Mei and Pope \cite{Lu:2008jk}. A natural question that we want to address in this paper is whether this geometry is linearly unstable and, if so, whether its instability teaches us something about the properties of the full geometry. 

These questions relating the stability properties of full geometries to those of their near-horizon geometries were analyzed in detail by Durkee and Reall  \cite{Durkee:2010ea}. They first observed that, in four and higher dimensions, any  known near-horizon geometry of Einstein gravity with a cosmological constant takes the form of a compact space ${\cal H}$ fibred over $AdS_2$. They further found that in all these near-horizon geometries, the study of linearized gravitational perturbations boils down to study a single Teukolsky-like master equation. The dependence of the perturbation on the compact space coordinates can be factored out by expanding the perturbation in eigenfunctions of a certain operator defined on ${\cal H}$. This effectively reduces the master equation to a form  that is precisely the one for the equation of a massive, charged, scalar field in $AdS_2$ with a homogeneous electric field (the latter being inherited from the rotation field of the full geometry). At this point,  one can define an ``effective Breitenl\"ohner-Freedman (BF) bound" for the scalar field, with the field being unstable if the effective mass of the field violates the bound. In this context, \cite{Durkee:2010ea} conjectured that instability of the near-horizon geometry does imply instability of the full black hole if the unstable mode respects certain symmetries and if appropriate boundary conditions are given. In 4 dimensions, the symmetry in question is axisymmetry. Supporting their conjecture, axisymmetric perturbations of NHEK do respect the BF bound, and the stability of such modes \cite{Amsel:2009ev,Dias:2009ex} is consistent with the stability of the full black hole. Further support for their conjecture comes from the near-horizon geometries of higher-dimensional Myers-Perry black holes. Axisymmetric instabilities of the near-horizon geometries were identified that precisely signal the onset of the axisymmetric ultraspinning instability  present in Myers-Perry black holes in $d\geq 6$ \cite{Dias:2010ma,Durkee:2010ea,Tanahashi:2012si}. Probably the only other system where the Durkee-Reall conjecture can be tested (using semi-analytical methods) is in the Kerr-AdS and NHEK-AdS pair of geometries since we just need to solve a Teukolsky master equation. Here, we will find that NHEK-AdS is stable against axisymmetric perturbations. This  is consistent with the stability of Kerr-AdS against axisymmetric perturbations and provides further support for the conjecture of \cite{Durkee:2010ea}. In addition, we do not find any instability in the non-axisymmetric sector of perturbations, {\it when} we impose outgoing boundary conditions at the asymptotic boundaries of NHEK-AdS (see discussion below).

A question that we leave open in our study is whether an analysis of perturbations in NHEK-AdS is able to capture a signature of the gravitational superradiant instability that is present in the full Kerr-AdS geometry \cite{Hawking:1999dp,Cardoso:2006wa}. At first glance the near-horizon geometry should be blind to this instability. The reason being that this instability requires the presence of two key ingredients, namely the existence of an ergoregion and of an asymptotic  reflecting wall. It is the multiple amplification/reflection that renders the system unstable. NHEK-AdS inherits the ergoregion from the full geometry but not its asymptotic boundary. Therefore, naively we would not expect to find a trace of an instability with a superradiant origin. However, the analysis might not be that simple and it could be the case that an appropriate choice of boundary conditions in NHEK-AdS is able to encode the reflecting boundary conditions of the full geometry. If this turns to be the case, our analysis misses it because we always impose outgoing boundary conditions.  
A detailed discussion of superradiant scattering in the near-NHEK-AdS geometry can be found in \cite{Chen:2010bh}. For similar reasoning, we cannot rule out the possibility that a different set of boundary conditions might lead to any other type of instability in NHEK-AdS.

Another question that we want to address concerns the Kerr/CFT correspondence originally formulated after an analysis of NHEK. This geometry has an $SL(2,R) \times
U(1)$ isometry group, where the $SL(2,R)$ extends the Kerr
time-translation symmetry and the $U(1)$ is simply inherited from the
axisymmetry of the Kerr solution. Guica, Hartman, Song and Strominger (GHSS)  conjectured that quantum gravity in the NHEK geometry with certain boundary conditions is equivalent to a chiral
conformal field theory (CFT) in 1+1 dimensions \cite{Guica:2008mu}. 
They then computed the microscopic entropy of the system and found it matches the Bekenstein-Hawking entropy of the associated extreme Kerr black hole.

The choice of boundary conditions plays a fundamental role in the analysis of \cite{Guica:2008mu} and is motivated entirely by considerations of the asymptotic symmetry group.
The GHSS ``fall-off" conditions specify  how the components $h_{\mu\nu}$  of the metric perturbations (about  the NHEK geometry) should behave asymptotically. GHSS's choice guarantees that the asymptotic symmetry group is generated by a time translation plus a single copy of the Virasoro algebra, the latter extending the $U(1)$ symmetry of the background.
However, as emphasized in \cite{Amsel:2009ev,Dias:2009ex}, NHEK (like AdS) is a non-globally hyperbolic spacetime. In other words, specifying initial data on a Cauchy surface is not enough to predict the future evolution of the system. This is because, in a Carter-Penrose diagram, these geometries have a timelike infinity that can be reached in finite time by null geodesics. Therefore to make classical predictions about  the future evolution of some initial data it is fundamental to specify also boundary conditions. Refs. \cite{Amsel:2009ev,Dias:2009ex} pointed out that we do not have the freedom to choose them arbitrarily. This is best illustrated if we consider a massive scalar field $\Phi$ in AdS$_d$. Solving the Klein-Gordon equation at the asymptotic boundary we find that the equation of motion selects the two only possible fall-offs  of the field, $\Phi \sim A \,r^{-\Delta_+}+ B\,r^{-\Delta_-}$. We are strictly restricted to select one of these decays and no other, if we want to preserve the asymptotic symmetry group.\footnote{One has $\Delta_\pm=\frac{d-1}{2}\pm  \sqrt{\frac{(d-1)^2}{4}+\mu^2\ell^2}$, where $\mu$ is the scalar field mass and $\ell$ the cosmological length. The requirement that the energy of the scalar field is finite further requires the scalar field mass to be above the Breitenl\"ohner-Freedman bound, and once it is above the unitarity bound,  only the mode with $r^{-\Delta_+}$ decay is normalizable.}
Similarly, the linearized Einstein equations in NHEK select the possible decays of the gravitational perturbations and \cite{Amsel:2009ev,Dias:2009ex} found  that these boundary conditions violate the GHSS ``fall-off" conditions.
The only exception are the axisymmetric modes (which furthermore do not excite non-axysymmetric modes at higher order in perturbation theory). 

The original Kerr/CFT correspondence has been extended to allow for a non-vanishing cosmological constant \cite{Lu:2008jk,Chen:2010bh} and to include higher-dimensional geometries (see \cite{Compere:2012jk} for a recent review). For these backgrounds, the original GHSS ``fall-off" conditions are still those required to have an asymptotic symmetry group generated by a time translation plus a single copy of the Virasoro algebra. So the fall-off is independent both of the  cosmological constant and of the spacetime dimension (the latter can be understood as consequence of the fact that the near-horizon geometry always contains an AdS$_2$ factor). 
Here, we  will look at the asymptotic behaviour of metric perturbations that solve the linearized Einstein equations in NHEK-AdS. The radial dependence of these perturbations can be found analytically and the desired perturbation decay is then obtained  through a simple series expansion. Like in the NHEK case, we find that these boundary conditions (except for the axisymmetric modes) violate the GHSS ``fall-off" conditions imposed in the Kerr/CFT formulation of \cite{Lu:2008jk}.
In higher dimensions, Ref. \cite{Godazgar:2011sn} recently determined the asymptotic behaviour of metric perturbations of the near-horizon geometry of the $d=5$ cohomogeneity-1 Myers-Perry black hole, where the problem can be addressed analytically.
Again, there are modes that violate the GHSS boundary conditions required in \cite{Guica:2010ej}.
The common conclusions of the present study in Kerr-AdS, together with \cite{Amsel:2009ev,Dias:2009ex,Godazgar:2011sn} in Kerr and higher dimensions, indicate that we still need to understand why  the Kerr/CFT ``fall-off" conditions and the boundary conditions required by classical physics to be predictable from initial data are different. Addressing this question would contribute to a deeper understanding of the correspondence. Recently, this question has started to be addressed in \cite{Bena:2012wc}, where it is found that there are  deformations of near-horizon geometries that obey the Kerr/CFT ``fall-off" conditions but are non-perturbative, i.e. they are not visible in a linear perturbative analysis of NHEK.

The plan of the paper is as follows. In Section \ref{sec:Kerr-AdS} we use the Teukolsky-Newman-Penrose formalism to find the decoupled master equation for arbitrary spin-$s$ perturbations in the Kerr-AdS black hole. This equation further separates into an angular equation, whose solutions are the AdS spin-weighted spheroidal harmonics, and into a radial equation. Section \ref{sec:pertNHEK} repeats the same exercise but this time in the NHEK-AdS geometry, which is the focus of our attention for the remainder of the paper. In Section \ref{sec:solutionAngRad} the eigenvalues of the angular equation are found numerically. On the other hand, the radial equation is solved exactly in terms of hypergeometric functions. In Section \ref{sec:PhysInterp} we look for linear instabilities in the NHEK-AdS geometry and we obtain, using the Hertz map, the asymptotic behaviour of the metric perturbations to compare them with the Kerr/CFT fall-off conditions. The physical interpretation and discussion of our findings are provided in this section. 
Appendix A provides a short summary of the Newman-Penrose formalism and the Teukolsky perturbation equations. In Appendix B we review the derivation of the the NHEK-AdS line element. Appendix C presents the master equation of the Kerr-AdS geometry in Poincar\'{e} coordinates.

\section{Master equation for perturbations of Kerr-AdS}\label{sec:Kerr-AdS}
We begin this section with a review of  properties of the Kerr-AdS spacetime relevant for our study. Subsequently we will present the Teukolsky master equation which governs perturbations around the Kerr-AdS background and we will separate it into a radial and an angular part. In the flat limit all results of this section exactly reproduce their counterparts in the Kerr geometry~\cite{Teukolsky:1973ha}.

\subsection{Properties of the spacetime}\label{sec:full geo_spacetime properties}
The Kerr-AdS geometry was found by Carter \cite{Carter:1968ks}. In the Boyer-Lindquist coordinate system $\{ \hat{t},\hat{r},\theta,\hat{\phi}\}$ it reads 
\begin{equation}
 ds^2
=-\frac{\Delta_r}{\Sigma^2}\left ( d\hat{t}-\frac{a}{\Xi}\sin^2\theta
\,d\hat{\phi}\right )^2 +\frac{\Sigma^2}{\Delta_r}\,d\hat{r}^2
+\frac{\Sigma^2}{\Delta_{\theta}}\,d\theta^2
   +\frac{\Delta_\theta}{\Sigma^2} \sin^2\theta\left (
a\,d\hat{t}-\frac{\hat{r}^2+a^2}{\Xi} \,d\hat{\phi}\right )^2 \,,
 \label{Kerr:metric}
\end{equation}
 where
 \begin{equation}
  \Delta_r=\left (\hat{r}^2+a^2\right )\left (1+\frac{\hat{r}^2}{\ell^2}
\right )-2M\hat{r}\,, \quad  \Xi=1-\frac{a^2}{\ell^2}\,, \quad
\Delta_{\theta}= 1-\frac{a^2}{\ell^2}\cos^2\theta\,, \quad
\Sigma^2=\hat{r}^2+a^2 \cos^2\theta  \,.
 \label{Kerr:metricAUX}
\end{equation}
This solution obeys $R_{\mu\nu} =-3\ell^{-2}g_{\mu\nu}$, and
asymptotically approaches AdS space with radius of curvature $\ell$.
The ADM mass and angular momentum of the black hole are $M/\Xi^2$
and $J=M a/\Xi^2$, respectively \cite{Gibbons:2004ai}. The event horizon is located at $\hat{r}=r_+$ (the
largest real root of $\Delta_r$). 

In this Boyer-Lindquist frame  the solution rotates asymptotically with angular velocity $\Omega_{\infty}=-a/\ell^2$. However, if we introduce the new coordinate system $\{ \hat{t},\hat{r},\theta,\hat{\varphi}\}=\{ \hat{t},\hat{r},\theta,\hat{\phi}+\frac{a}{\ell ^2} \hat{t}\}$ we get the Kerr-AdS solution written in a non-rotating frame at infinity. The horizon
angular velocity measured with respect to this non-rotating frame at infinity is
\begin{equation} \label{kerr:OmegaH}
\Omega_H=\frac{a}{r_+^2+a^2}\left ( 1+\frac{r_+^2}{\ell^2} \right
)\,.
\end{equation}
This is the angular velocity that is relevant for the thermodynamic analysis of the Kerr-AdS black hole \cite{Caldarelli:1999xj,Hawking:1999dp,Gibbons:2004ai,Caldarelli:2008ze}. Henceforth we will work in the non-rotating frame.

The rotation parameter is bounded by $a<\ell$.
 Solutions saturating this bound do not describe black holes. In the limit $a \rightarrow \ell$ at fixed $r_+$, the mass and angular momentum of the black hole diverge, and the circumference of the black hole as measured at the equator becomes infinitely large in this limit. The temperature is given by
\begin{equation}  \label{kerr:T}
 T_H=\frac{r_+}{2 \pi
}\left(1+\frac{r_+^2}{\ell ^2}\right)\frac{1}{r_+^2+a^2}-\frac{1}{4
\pi r_+}\left(1-\frac{r_+^2}{\ell ^2}\right)\,.
\end{equation}

The Kerr-AdS black hole has a regular extremal configuration where its temperature vanishes while the entropy remains finite. The extremality conditions $T_H=0$ and $\Delta_r(r_+)=0$ allow us to express $a=a_{\rm ext}$ and $M=M_{\rm ext}$ as functions of $\ell$ and $r_+$,
\begin{equation}\label{full geo_extremality relations}
a_{\rm ext}= r_+ \sqrt{\frac{3 r_+^2+\ell
^2}{\ell^2-r_+^2}} \,,\qquad M_{\rm ext}=\frac{r_+\left(1+r_+^2\ell^{-2}\right)^2}{1-r_+^2\ell^{-2}}\,.
\end{equation}
At extremality, we further have $\Omega_H=\Omega_H^{\rm ext}$ with
 \begin{equation} \label{kerr:extreme} 
 \Omega_H^{\rm
ext}=\frac{\sqrt{\ell^4+2 r_+^2 \ell^2-3 r_+^4}}{2 r_+ \ell^2}\,,
\qquad \hbox{and} \quad \frac{r_+}{\ell}<\frac{1}{\sqrt{3}}\,.
 \end{equation}
 Note that only black holes with $r_+/\ell<3^{-1/2}$ can reach zero temperature by virtue of the constraint $a<\ell$. Some further properties of the Kerr-AdS spacetime are discussed in Appendix A of~\cite{Dias:2010ma}.
\subsection{Master equation}\label{sec:full geo_master eq}

Teukolsky investigated perturbations of the Kerr geometry \cite{Teukolsky:1973ha} using the  the Newman-Penrose formalism. To be self-contained, we briefly review this formalism and Teukolsky's master equation in Appendix \ref{app:NP & Teukolsky}. This master equation holds for any Petrov type D background, and thus, in particular, it governs perturbations in the Kerr-AdS black hole.  In his original analysis, Teukolsky 
makes use of an affinely parametrized null tetrad $-$ the outgoing Kinnersly tetrad $-$ that is regular in the past horizon \cite{Teukolsky:1974yv}. To guarantee that the flat limit of our calculations exactly reproduces Teukolsky's results, we choose to work with the natural extension of Kinnersly's tetrad to AdS. Other choices are possible; in particular perturbations of the Kerr-AdS geometry have previously been studied in~\cite{Chambers:1994ap,Giammatteo:2005vu} using a tetrad that is not affinely parametrized (but that suits the symmetries of the problem), and in the rotating Boyer-Lindquist frame.

To find the Teukolsky master equation for spin-$s$ perturbations in the Kerr-AdS geometry we work with the Newman-Penrose (NP) null tetrad ${\bf e}_{a}=\{\bm{\ell},\bm{n},\bm{m},\bm{\overline{m}}\}$ (the bar demotes complex conjugation),
\begin{eqnarray}
 &&\NPl^{\mu}\partial_{\mu}=\frac{1}{\Delta _r}\left(  \left(\hat{r}^2+a^2\right)\partial_{\hat{t}}+\Delta _r \partial_{\hat{r}}+a\left(1+\frac{ \hat{r}^2}{\ell ^2}\right)\partial_{\hat{\varphi}} \right) ,\nonumber\\
 && \NPn^{\mu}\partial_{\mu}=\frac{1}{2\Sigma^2 } \left( \left(\hat{r}^2+a^2\right)\partial_{\hat{t}}-\Delta _r\partial_{\hat{r}}+a\left(1+\frac{ \hat{r}^2}{\ell ^2}\right)\partial_{\hat{\varphi}}  \right),\nonumber\\
 &&  \NPm^{\mu}\partial_{\mu}= \frac{\sin  \theta }{\sqrt{2}\sqrt{\Delta _{\theta }}(\hat{r}+i a \cos  \theta )}\left(i \,a\, \partial_{\hat{t}}+\frac{\Delta _{\theta }}{\sin  \theta }\,\partial_{\theta}+\frac{i \Delta _{\theta }}{\sin ^2\theta }\,\partial_{\hat{\varphi}} \right).
 \label{KerrAdS:NPtetrad}
\end{eqnarray}
Using this null basis we can construct the NP spin coefficients, the complex Weyl scalars and the NP directional derivative operators. A brief, but self-contained, review of the NP formalism is given in Appendix  \ref{app:NP & Teukolsky}.
The Kerr-AdS  black hole is a Petrov type D background since the only non-vanishing complex Weyl scalar is 
$\Psi_2 = - M (r- i a\cos\theta)^{-3}$.
The perturbations of spin-$s$ fields in a type D background are described by the Teukolsky  decoupled  equations, namely by equations (2.12)-(2.15),  (3.5)-(3.8), and (B4)-(B5) of \cite{Teukolsky:1973ha}. 
We collect these equations in a compact form in the pair of equations \eqref{Teukolskyeqs_pos spin} and \eqref{Teukolskyeqs_neg spin}  of Appendix  \ref{app:NP & Teukolsky}.
In the following discussion spin $s=\pm 2, \pm 1,\pm 1/2, \pm 3/2, 0$ describes, respectively, gravitational, electromagnetic, fermionic, and massless uncharged scalar field perturbations.

Inserting the NP quantities constructed out of the null basis  \eqref{KerrAdS:NPtetrad} into the Teukolsky equations \eqref{Teukolskyeqs_pos spin} and \eqref{Teukolskyeqs_neg spin},  we get the Teukolsky master equation for spin $s={\pm 2, \pm 1, \pm 3/2, \pm 1/2}$ in the Kerr-AdS background,
\begin{align}\label{KerrAdS:TeukEq}
&\left[\frac{\left(\hat{r}^2+a^2\right)^2}{\Delta _{\hat{r}} }-\frac{a^2 \sin^2\theta}{\Delta _{\theta}}\right] \partial_{\hat{t}}^{\,2}\Psi ^{(s)}+2a \left[\frac{ \left(\hat{r}^2+a^2\right) \left(\hat{r}^2+\ell ^2\right)}{\ell ^2\Delta _{\hat{r}}}-1 \right] \partial_{\hat{t}}\partial_{\hat{\varphi} }\Psi ^{(s)} \nonumber\\
&+\left[\frac{ a^2 \left(\hat{r}^2+\ell ^2\right)^2 }{\ell^4 \Delta _{\hat{r}} }-\frac{\Delta _{\theta}}{\sin^2\theta}\right] \partial _{\hat{\varphi} }^{\,2}\Psi ^{(s)}-\Delta _{\hat{r}}^{-s}\partial_{\hat{r}}\left(\Delta _{\hat{r}}^{s+1}\partial_{\hat{r}} \Psi ^{(s)}\right)\nonumber\\
&-\frac{1}{\sin\theta}\partial _{\theta}\left(\sin\theta\,\Delta _{\theta}\,\partial _{\theta} \Psi ^{(s)}\right) +s\left[\frac{4\hat{r} \Delta _{\hat{r}}-\left(\hat{r}^2+a^2\right) \Delta _{\hat{r}}'}{\Delta _{\hat{r}}}+i\,\frac{2a \,\Xi \, \cos\theta}{\Delta _{\theta} }\right] \partial_{\hat{t}}\Psi ^{(s)}\nonumber\\
&-\frac{s}{\ell ^2}\left[\frac{ a \left(\hat{r}^2+\ell ^2\right) \Delta _{\hat{r}}'}{\Delta _{\hat{r}}}-4 a \hat{r}+i\,\frac{2\,\ell ^2\,\Xi \, \cos\theta}{\sin^2\theta} \right] \partial _{\hat{\varphi} }\Psi ^{(s)} +\Biggl\{\left(16\,s^8-120\,s^6+273\,s^4\right)\frac{\Sigma^2}{18\,\ell^2}\nonumber\\
&+s^2\left[\frac{\Xi}{\sin^2\theta}-\frac{\Xi}{\Delta_{\theta}}-\frac{\left(277\,\hat{r}^2+205\,a^2\cos^2\theta\right)}{18\,\ell^2}\right]-s\left(1+\frac{a^2}{\ell^2}+\frac{6\hat{r}^2}{\ell^2}\right)\Biggr\} \Psi ^{(s)}=4\pi \mathcal{T}_{(s)}\,,
\end{align}
where we have allowed for a possible source term  $\mathcal{T}_{(s)}$ on the right hand side and $\Delta _{\hat{r}}'=\partial_r \Delta _{\hat{r}}$. Setting $s=0$ in this master equation we get the Klein-Gordon equation for a massless scalar field. 

The relation  between the master fields $\Psi^{(s)}$ and the perturbed Weyl scalars (that we represent using the notation $\delta Q$) is 
\begin{alignat}{4}\label{NHgeometry_redef scalars}
&\Psi^{(2)}=\delta \Psi_0\,, &\quad &\Psi^{(1)}=\delta\phi_0\,, &\quad&\Psi^{(\frac{1}{2})}=\delta\chi_0\,, 
&\quad& \Psi^{(\frac{3}{2})}=\,\delta \Phi_0\,,
\nonumber \\
&\Psi^{(-2)}=(-\Psi_2)^{-\frac{4}{3}}\,\delta\Psi_4\,, &\quad &\Psi^{(-1)}=(-\Psi_2)^{-\frac{2}{3}}\,\delta\phi_2\,, &\quad&\Psi^{(-\frac{1}{2})}=(-\Psi_2)^{-\frac{1}{3}}\,\delta\chi_1\,,
 &\quad& \Psi^{(-\frac{3}{2})}=\,(-\Psi_2)^{-1}\,\delta\Phi_3\,.
\end{alignat}
The fields $\delta\Psi_0$, $\delta\Psi_4$ and $\delta\phi_0$, $\delta\phi_2$ are the perturbations of the usual Weyl and Maxwell scalars of the Newman-Penrose formalism (see Appendix~\ref{app:NP & Teukolsky} for details), while $\delta\chi_0$, $\delta\chi_1$ are the components of the neutrino spinor and $\delta\Phi_0$, $\delta\Phi_3$ are the components of the Rarita-Schwinger field. 
Likewise the master equation source terms $\mathcal{T}_{(s)}$ are defined via
\begin{alignat}{4}\label{NHgeometry_redef sources}
&\mathcal{T}_{(2)}=\mathcal{T}_{\Psi_0}\,, &\quad &\mathcal{T}_{(1)}={\textstyle \frac{1}{2}}\mathcal{T}_{\phi_0}\,, &\quad&\mathcal{T}_{(\frac{1}{2})}={\textstyle \frac{1}{4}}\mathcal{T}_{\chi_0}\,,  &\quad &
\mathcal{T}_{(3/2)}=\,{\textstyle \frac{3}{4}}\mathcal{T}_{\Phi_0}\,,
\nonumber \\
&\mathcal{T}_{(-2)}=(-\Psi_2)^{-\frac{4}{3}}\,\mathcal{T}_{\Psi_4}\,, &\quad &\mathcal{T}_{(-1)}={\textstyle \frac{1}{2}}(-\Psi_2)^{-\frac{2}{3}}\,\mathcal{T}_{\phi_2}\,, &\quad&\mathcal{T}_{(-\frac{1}{2})}={\textstyle \frac{1}{4}}(-\Psi_2)^{-\frac{1}{3}}\,\mathcal{T}_{\chi_1}\,, &\quad&\mathcal{T}_{(-3/2)}=\,{\textstyle \frac{3}{4}}(-\Psi_2)^{-1}\,\mathcal{T}_{\Phi_3}\,,
\end{alignat}
where the source terms  $\{ T_0\equiv\mathcal{T}_{\Psi_0}, T_4\equiv\mathcal{T}_{\Psi_4}, J_0\equiv \mathcal{T}_{\phi_0}, J_2\equiv \mathcal{T}_{\phi_2}, T_{\chi_0}, T_{\chi_1} \}$  can be found in equations (2.13), (2.15), (3.6) and (3.8) of Appendix B of \cite{Teukolsky:1973ha}. 

Onwards let us restrict our attention to the AdS vacuum case where no sources are present, $\mathcal{T}_{(s)}\equiv 0$.
Introducing the  separation constant $\hat{\Lambda}_{lm\hat{\omega}}^{(s)}$ and the {\it ansatz}
\begin{equation}\label{ansatzWeyl}
\Psi^{(s)}=e^{-i \,\hat{\omega} \,\hat{t}}e^{i\,m \hat{\varphi}} \,\Phi_{lm \hat{\omega}}^{(s)}(\hat{r})  S^{(s)}_{lm \hat{\omega}}(\theta)\,,
\end{equation}
the Teukolsky master equation separates.
The radial equation is
\begin{equation}\label{KerrAdS:radialEq}
\hspace{-4cm}\Delta _r^{-s}\partial_{\hat{r}}\left[\Delta _r^{s+1}\partial_{\hat{r}}\Phi_{lm \hat{\omega} }^{(s)}(\hat{r})\right]+H(\hat{r}) \,\Phi _{lm \hat{\omega} }^{(s)}(\hat{r})=0\,,
\end{equation}
with
\begin{eqnarray}\label{KerrAdS:radialEqAUX}
&&H(\hat{r})=\frac{K_T^2- i\, s\,  \Delta _r' K_T}{\Delta _r }+2 \, i\,  s \,  K_T'+ \frac{s+|s|}{2}\Delta _r''  \\
&&\hspace{1.4cm}
 -|s|\left(|s|-1\right)\left(2|s|-1\right)\left(2|s|-7\right)\frac{\hat{r}^2}{3\ell^2}-|s|\left(|s|-2\right)\left(4s^2-12|s|+11\right)\frac{a^2}{3\ell^2}-\hat{\lambda}_{lm\hat{\omega} }^{(s)}\,, \nonumber \\
&&K_T(\hat{r})=\hat{\omega} \left(\hat{r}^2+a^2\right)- m\, a \left(1+\frac{\hat{r}^2}{\ell ^2}\right),   \quad \hbox{and} \quad \hat{\lambda}_{lm \hat{\omega} }^{(s)}=\hat{\Lambda}_{lm\hat{\omega}}^{(s)}-2\, a \,m \,\hat{\omega} +a^2 \hat{\omega} ^2+(s+|s|)\,,\nonumber
\end{eqnarray}
while the angular equation reads
\begin{equation}\label{KerrAdS:angEq}
\begin{split}
&\frac{1}{\sin  \theta }\,\partial_{\theta }\left(\sin  \theta \, \Delta _{\theta }\,\partial _{\theta } S_{lm \hat{\omega} }^{(s)}(\theta )\right) +\biggl[(a \,\hat{\omega} \, \cos  \theta)^2\frac{\Xi }{\Delta _{\theta }}-2\,s \,a \,\hat{\omega} \, \cos  \theta \frac{\Xi }{\Delta _{\theta}}+s+\hat{\Lambda}_{lm\hat{\omega}}^{(s)} \\
& \hspace{5.3cm}-  \left(m+s \,\cos \theta \,\frac{ \Xi }{\Delta _{\theta }}\right)^2 \frac{\Delta _{\theta }}{\sin ^2\theta }-2\delta_s\frac{a^2}{\ell ^2} \sin ^2\theta \biggr] S_{lm \hat{\omega} }^{(s)}(\theta )=0\,,
\end{split}
\end{equation}
with $\delta_s=1$ if $|s|=\{2,1,1/2, 3/2 \}$ and $\delta _s=0$ if $s=0$.
Note that in the limit $\ell\to\infty$,  equations \eqref{KerrAdS:TeukEq}, \eqref{KerrAdS:radialEq} and \eqref{KerrAdS:angEq} reduce to the standard Teukolsky equations for the asymptotically flat Kerr background.

As usual when separating variables we are free to move a constant from the radial to the angular equation. We tuned the constant terms in equation~\eqref{KerrAdS:angEq} such that its flat limit precisely agrees with
\begin{equation}
 \frac{1}{\sin\theta}\,\frac{d}{d\theta}\left( \sin\theta \frac{d}{d\theta}
S_{lm\hat{\omega}}^{(s)}(\theta)\right)
 +\left[ (C\,\cos\theta)^2-2 s\, C \,\cos\theta +s +\hat{\Lambda}_{lm\hat{\omega}}^{(s)} -\frac{\left(
m+s\,\cos\theta \right)^2}{\sin^2\theta}\right]S_{lm\hat{\omega}}^{(s)}(\theta)=0\,,
 \label{AngEq}
\end{equation}
with $C=a \hat{\omega}$ which is the standard form of the spin-weighted  spheroidal harmonic equation \cite{Breuer1977,Berti:2005gp}.
Equation \eqref{KerrAdS:angEq} is the natural extension of \eqref{AngEq} when the cosmological constant is switched-on. Hence its eigenfunctions can naturally be called the spin-weighted {\it AdS} spheroidal harmonics, $e^{i m\hat{\varphi}}S_{lm\hat{\omega}}^{(s)}(\theta)$, with positive integer $l$ specifying the
number of zeros, $l-{\rm max}\{|m|,|s|\}$, along the polar direction of the eigenfunction.
The associated eigenvalues $\hat{\Lambda}_{lm\hat{\omega}}^{(s)}$ can be computed
numerically. They are a function of $s,l,m$ and regularity imposes the constraints that $-l\leq m\leq l$ must be an integer and $l\geq |s|$. To leading order in $a/\ell$ (note that $a/\ell \ll 1$ implies $a\hat{\omega} \ll 1$),  one has
$\hat{\Lambda}_{lm\hat{\omega}}^{(s)}=(l-s)(l+s+1)+\mathcal{O}(a/\ell)$, i.e. at this order the eigenvalues of  \eqref{KerrAdS:angEq} reduce to those of the well known spin-$s$ spherical harmonic equation.

In the flat space limit, $\ell\to \infty$, when the black hole is extremal and the perturbations have a frequency that saturates the superradiant bound, i.e. $\hat{\omega}=m \Omega_H^{\rm ext}$, \eqref{KerrAdS:radialEq} reduces to a hypergeometric equation and thus has an exact solution in terms of hypergeometric functions. This was first observed in  \cite{Teukolsky:1974yv}. However, for non-vanishing cosmological constant we can no longer solve the radial equation analytically even in the above particular case.
Finally, note that the radial and angular equations also describe perturbations of Kerr-de Sitter black holes if we do the trade $\ell^2\to -\ell^2$ (see also \cite{Suzuki:1999nn}). 

\section{Master equation for perturbations of NHEK-AdS }\label{sec:pertNHEK}

In this section, we first briefly discuss some properties of the NHEK-AdS geometry. Then we obtain the associated master equation which governs its perturbations and separate it into a radial and an angular part. In the flat limit $\ell\to\infty$ all our results agree with their counterparts of the NHEK geometry~\cite{Dias:2009ex}.

\subsection{Properties of the spacetime}\label{sec:NH geo_spacetime properties}
The Kerr-AdS black hole has an extreme regular configuration where its temperature vanishes but the entropy remains finite. We can then take the near-horizon limit of this extreme Kerr-AdS black hole, and  get the  Kerr-AdS near-horizon geometry (NHEK-AdS), as done in  \cite{Lu:2008jk}. This limit is reviewed in Appendix \ref{sec:AppendixNHgeometry}: we start with the coordinates $\{ \hat{t},\hat{r},\theta,\hat{\phi}\}$ of \eqref{Kerr:metric} and we end up with the near-horizon coordinates $\{t,r,\theta,\phi\}$.
The NHEK-AdS gravitational field then reads \cite{Lu:2008jk}
\begin{equation}\label{NHgeometry_NH line element}
ds^2=\frac{\Sigma_+^2}{V}\Biggl[-\left(1+r^2 \right)dt^2+\frac{dr^2}{1+r^2}+\frac{Vd\theta^2}{\Delta_{\theta}}\Biggr]+\frac{\sin^2\theta\Delta_{\theta}}{\Sigma_+^2}\frac{(r_+^2+a^2)^2}{\Xi^2}\Biggl(d\phi+\frac{2a\,r_+ \Xi}{V(r_+^2+a^2)}\,rdt\Biggr)^2\,,
\end{equation}
with $\Delta_{\theta}(\theta)$ and $\Xi$ defined in \eqref{Kerr:metricAUX}, and 
\begin{alignat}{1}
\Sigma_+^2=&r_+^2+a^2\cos^2\theta\,, \qquad V=\frac{1+6r_+^2\ell^{-2}-3r_+^4\ell^{-4}}{1-r_+^2\ell^{-2}}\,,
\end{alignat}
and it obeys $R_{\mu\nu} =-3\ell^{-2}g_{\mu\nu}$. The rotation parameter $a$ is constrained to obey $a<\ell$ and the extremality condition \eqref{kerr:extreme}, i.e.
 \begin{equation} 
\label{NHgeometry_extremality condition}
a= r_+ \sqrt{\frac{3 r_+^2+\ell
^2}{\ell^2-r_+^2}} \qquad \hbox{and} \quad a<\ell \quad \Rightarrow \quad\frac{r_+}{\ell}<\frac{1}{\sqrt{3}}\,. \end{equation}
Onwards, although we will keep the parameter $a$ in our results for the benefit of compactness, the reader should keep in mind that it is not an independent parameter and that the constraint \eqref{NHgeometry_extremality condition} is implicit.

NHEK-AdS has the property that surfaces of constant $\theta$ are warped $AdS_3$ geometries, i.e. a
circle fibred over $AdS_2$ with warping parameter proportional to $g_{\phi\phi}$. 
The isometry group is $SL(2,R) \times U(1)$. Quite importantly, NHEK-AdS is a non-globally hyperbolic spacetime, having timelike infinities both at $r =-\infty$ and $r=+\infty$.  It also has an ergoregion (where
the Killing field $\partial/\partial t$ is spacelike) which extends to $r=\pm \infty$.

\subsection{Master equation}\label{sec:NH geo_master eq}
We are interested in linear perturbations of the NHEK-AdS geometry.
To obtain the associated  Teukolsky master equation for spin-$s$ perturbations, we work
with the Newman-Penrose null tetrad basis
\begin{eqnarray}\label{NHEKAdS:NPbasis}
&& \NPl^{\mu}\partial_{\mu}=\frac{V}{1+r^2}\biggl(\partial_t+\left(1+r^2\right)\partial_r-\frac{2 a r r_+ \Xi}{\left(r_+^2+a^2\right)V}\partial_{\phi}\biggr)\,,\nonumber \\
&& \NPn^{\mu}\partial_{\mu}=\frac{1}{2\Sigma_+^2}\biggl(\partial_t-\left(1+r^2\right)\partial_r-\frac{2 a r r_+ \Xi}{\left(r_+^2+a^2\right)V}\partial_{\phi}\biggr)\,,\nonumber \\
&& \NPm^{\mu}\partial_{\mu}=\frac{\sqrt{\Delta_{\theta}}}{\sqrt{2}\left(r_++i a \cos\theta\right)}\biggl(\partial_{\theta}+i\frac{\Xi \,\Sigma_+^2}{\left(r_+^2+a^2\right)\sin\theta\Delta_{\theta}}\partial_{\phi}\biggr)\,.
\end{eqnarray}
The  NP spin coefficients can be obtained from this tetrad and \eqref{spincoef}. The non-vanishing ones are
\begin{alignat}{3}\label{NHEKAdS:spinCoef}
\alpha=&-\frac{r_+\cos\theta\bigl(\ell^2+a^2\left(1-2\cos^2\theta\right)\bigr)-ia\bigl(\ell^2\left(2-\cos^2\theta\right)-a^2\cos^2\theta\bigr)}{2\sqrt{2}\,\ell^2\left(r_+\!-i a\cos\theta\right)^2\sin\theta\sqrt{\Delta_{\theta}}}\,, \qquad  
\gamma=\frac{r}{2\Sigma_+^2}\,,\nonumber \\
\beta=&\frac{\cos\theta\bigl(\ell^2+a^2(1-2\cos^2\theta)\bigr)}{2\sqrt{2}\,\ell^2\left(r_+\!+i a\cos\theta\right)\sin\theta\sqrt{\Delta_{\theta}}}\,, \qquad 
\pi=\frac{i\,a\sin\theta\sqrt{\Delta_{\theta}}}{\sqrt{2}\left(r_+\!-i a \cos\theta\right)^2}\,,  \qquad  
\tau= -\frac{i \,a\sin\theta\sqrt{\Delta_{\theta}}}{\sqrt{2}\Sigma_+^2}\,.
\end{alignat}
NHEK-AdS is a Petrov type D geometry since the  only non-vanishing complex Weyl scalar is 
\begin{equation}\label{NHEKAdS:Psi2}
\Psi_2 =-\frac{(a^2+r_+^2)^2}{r_+(a^2+3r_+^2)(r_+-i a\cos\theta)^3}\,.
\end{equation}

Inserting the NP spin coefficients \eqref{NHEKAdS:spinCoef}  and the directional derivatives associated with the basis  \eqref{NHEKAdS:NPbasis} into the Teukolsky equations \eqref{Teukolskyeqs_pos spin} and \eqref{Teukolskyeqs_neg spin},  we get the Teukolsky master equation for spin $s={\pm 2, \pm 1, \pm 3/2, \pm 1/2}$ in the NHEK-AdS background,
\begin{alignat}{2}\label{NHgeometry_master eq}
&\frac{V}{1+r^2}\,\partial_t^{\,2}\Psi^{(s)}
-\frac{4a\,r\,r_+\,\Xi}{\left(1+r^2\right)\left(r_+^2+a^2\right)}\,\partial_t\partial_{\phi}\Psi^{(s)}+\frac{\Xi^2}{{\left(r_+^2+a^2\right)}^2}\Biggl[\frac{4a^2r^2r_+^2}{V\left(1+r^2\right)}\nonumber\\
&-\frac{\Sigma_+^{\,4}}{\Delta_{\theta}\sin^2\theta}\Biggr]\partial_{\phi}^{\,2}\Psi^{(s)}-V\left(1+r^2\right)^{-s}\partial_r\left[ \left(1+r^2\right)^{s+1}\,\partial_r\Psi^{(s)}\right]\nonumber\\
&-\frac{1}{\sin\theta}\,\partial_{\theta}\Bigl(\sin\theta\Delta_{\theta}\,\partial_{\theta}\Psi^{(s)}\Bigr)-2s\,\frac{V\,r}{1+r^2}\,\partial_t\Psi^{(s)}-2s\,\frac{\Xi}{r_+^2+a^2}\Biggl[\frac{2a\,r_+}{1+r^2}\nonumber\\
&+i\cos\theta\left(\frac{a^2\left(r_+^2+\ell^2\right)}{\ell^2\Delta_{\theta}}+\frac{r_+^2+a^2}{\sin^2\theta}\right)\Biggr]\partial_{\phi}\Psi^{(s)}+\Biggl\{\left(16\,s^8-120\,s^6+273\,s^4\right)\frac{\Sigma_+^{\,2}}{18\ell^2}\nonumber\\
&+s^2\left[\frac{\Xi}{\sin^2\theta}-\frac{\Xi}{\Delta_{\theta}}-\frac{\left(277\,r_+^2+205\,a^2\cos^2\theta\right)}{18\ell^2}\right]-sV\Biggr\}\Psi^{(s)}=4\pi \mathcal{T}_{(s)}\,.
\end{alignat}
The relation  between the master fields $\Psi^{(s)}$ and the perturbed Weyl scalars is given by  \eqref{NHgeometry_redef scalars},  with the background $\Psi_2$ defined in \eqref{NHEKAdS:Psi2}, and the master source terms $\mathcal{T}_{(s)}$ are defined via \eqref{NHgeometry_redef sources}. Setting $s=0$ in this master equation we get the Klein-Gordon equation for a massless scalar field in the NHEK-AdS geometry. 

To solve the above master equation we introduce the separation constant $\Lambda^{(s)}_{lm}$, and make the  separation ansatz
\begin{equation}\label{NHgeometry_separation ansatz}
\Psi^{(s)}=e^{-i\omega t}e^{im\phi}\left(1+r^2\right)^{-s/2}\Phi^{(s)}_{lm\omega}(r)S^{(s)}_{lm}(\theta)\,.
\end{equation}
The resulting angular equation is (with $a$ given by \eqref{NHgeometry_extremality condition})
\begin{fleqn}
\begin{align}\label{NHgeometry_angular master eq}
&\frac{1}{\sin\theta}\,\partial_\theta\left(\!\sin\theta\Delta_{\theta}\partial_\theta\,S^{(s)}_{lm}(\theta)\!\right)+\Biggl[-\left(16\,s^8-120\,s^6+273\,s^4\right)\frac{\Sigma_+^{\,2}}{18\ell^2}+s^2\,\biggl(-\frac{a^2\cos^2\theta\,\Xi}{\ell^2\Delta_{\theta}}-\frac{\Xi}{\sin^2\theta}\nonumber\\
&+\frac{\Xi}{\,\,\,\Delta_{\theta}}+\frac{\left(277\,r_+^2+205\,a^2\cos^2\theta\right)}{18\ell^2}\biggr)-\frac{m\left(m+2s\cos\theta\right)\Xi}{\sin^2\theta}
+\frac{16\,m^2a^4}{\ell^2\left(r_+^2-a^2\right)}+\frac{16m^2a^6\left(a^2+7r_+^2\right)}{V\ell^4{\left(r_+^2-a^2\right)}^2}\Biggr) \nonumber\\
&+\frac{\Xi }{\ell^2 \Delta _{\theta}}\left(\frac{2 a \,m \,r_+^2}{\left(r_+^2-a^2\right)}+s\, a \cos \theta\right)^2+V\biggr(\frac{a^2m^2}{\left(r_+^2-a^2\right)}+\underbrace{s-\frac{7m^2}{4}+\Lambda^{(s)}_{lm}}_{-s^2+\tLambda^{(s)}_{lm}}\biggl) \,\Biggr]S^{(s)}_{lm}(\theta)=0\,.
\end{align}
\end{fleqn}
while the radial equation for $\Phi^{(s)}_{lm\omega}(r)$ reads
\begin{equation}\label{NHgeometry_radial master eq}
\frac{d}{dr}\biggl[\left(1+r^2\right)\frac{d}{dr}\Phi^{(s)}_{lm\omega}(r)\biggr]-\biggl[\mu^2-\frac{\left(\omega+qr\right)^2}{1+r^2}\biggr]\Phi^{(s)}_{lm\omega}(r)=0\,,
\end{equation}
where
\begin{equation}\label{def:muq}
\mu^2=q^2+s^2+s-\frac{7m^2}{4}+\Lambda^{(s)}_{lm}\equiv q^2+\tLambda^{(s)}_{lm}\,,\qquad q=\frac{2a\,m\,r_+\Xi}{\left(r_+^2+a^2\right)V}-i\,s\,.
\end{equation}
We have introduced the shifted eigenvalues $\tLambda^{(s)}_{lm}$  which have the  advantage of having  the symmetry $\tLambda^{(s)}_{lm}=\tLambda^{(-s)}_{lm}$ (since $\Lambda _{lm}^{(-s)}=\Lambda _{lm}^{(s)}+2s$) that will be useful later. This follows from the property $S^{(s)}_{lm}(\pi-\theta)=S^{(s)}_{lm}(\theta)$. Moreover, when $\ell \to \infty$ one has $q\to m-i\,s$, in agreement with the asymptotically flat limit  result \cite{Dias:2009ex}.

An interesting observation, first made in the NHEK case  \cite{Dias:2009ex},  that also holds in NHEK-AdS,  is that  the radial equation \eqref{NHgeometry_radial master eq} is exactly the equation for a scalar field of mass $\mu$ and  charge $q$, 
\begin{equation}\label{KGads2}
\left(\mathcal{D}^2+\mu^2\right)\Phi=0\,,\qquad \mathcal{D}_{\mu}=\nabla_{\mu}-iqA_{\mu}\,,
\end{equation}
in $AdS_2$ space with curvature radius $\ell_{AdS_2}=1$ and with an electric field, 
\begin{equation}
ds^2=\left(1+r^2\right)dt^2-\frac{1}{\left(1+r^2\right)}dr^2\,,   \qquad A=rdt\,.
\end{equation}
Indeed, if we introduce the  separation ansatz $\Phi(r,t)=e^{-i\,\omega\, t} \Phi^{(s)}_{lm\omega}(r)$ into  \eqref{KGads2}, the Klein-Gordon equation exactly reproduces equation \eqref{NHgeometry_radial master eq}. 
Therefore a general spin-$s$ perturbation with angular momentum $m$ in NHEK-AdS
obeys  the wave equation for a massive charged scalar field in
$AdS_2$ with a homogeneous electric field. Interestingly, the charge $q$ and squared mass $\mu^2$ are complex, although $\mu^2-q^2$ is real. A massive charge scalar field in $AdS_2$ with homogeneous electric field was first studied in Ref. \cite{Strominger:1998yg}, and our radial solutions will necessarily reproduce those found in \cite{Strominger:1998yg}.

An intriguing property of the angular equation \eqref{NHgeometry_angular master eq} is that it does not depend on the frequency of the perturbation, contrary  to what happens in the Kerr-AdS angular equation \eqref{KerrAdS:angEq}. This property is best understood if we analyze what happens to the perturbation frequency in the near-horizon limit procedure. For simplicity consider the near-horizon transformation, reviewed in Appendix \ref{sec:AppendixNHgeometry}, that takes the Kerr-AdS geometry in the frame $\{ \hat{t}, \hat{r},\theta,\hat{\varphi}\}$ into NHEK-AdS in Poincar\'e coordinates $\{ t', r',\theta,\phi'\}$ written in \eqref{NHEKpoincare}. In this process a Kerr-AdS mode with frequency $\hat{\omega}$ and azimuthal quantum number $m$ transforms as
\begin{equation}
e^{i \,m \,\hat{\varphi}-i \,\hat{\omega}\, \hat{t}}\to e^{i\, m\, \phi'-i \frac{1}{\lambda} \frac{r_+^2+a^2}{V r_+}\left( \hat{\omega}-m\Omega_H^{\rm ext} \right) t'}\equiv e^{i m \phi'-i \omega' t'}\,,
\end{equation}
that is, the Kerr-AdS frequency $\hat{\omega}$  is related to  the NHEK-AdS frequency $\omega'$ by 
\begin{equation}\label{relationFreq}
\frac{r_+^2+a^2}{V r_+}\left( \hat{\omega}-m\Omega_H^{\rm ext} \right) = \lim_{\lambda\to 0} \lambda \,\omega'\,,
\end{equation}
where $\lambda\to 0$ is the quantity that zooms the near-horizon region of the original black hole (see Appendix \ref{sec:AppendixNHgeometry}). We conclude that {\it all} finite frequencies $\omega'$ in the NHEK-AdS throat correspond to the {\it single} frequency $\hat{\omega}=m\Omega_H^{\rm ext}$ in the extreme Kerr-AdS black hole
(this property was first observed by \cite{Bardeen:1999px} in the Kerr case).
Moreover, the frequency $\hat{\omega}=m\Omega_H^{\rm ext}$  is exactly the one that saturates the superradiant bound of extreme Kerr-AdS.

\section{Solution of the radial and angular equations in NHEK-AdS }\label{sec:solutionAngRad}

In this section we find the solutions of the radial equation \eqref{NHgeometry_radial master eq}  and of the angular equation \eqref{NHgeometry_angular master eq}. The radial equation can be solved exactly in terms of hypergeometric functions. The angular equation can be solved numerically with very high accuracy. Since it is independent of the frequency, we can solve it independently of the radial equation solution. Once its eigenvalues are found we insert them in the radial solution to study the physical properties of the system. In the flat limit $\ell\rightarrow\infty$ our results reduce to those found in the analysis of NHEK \cite{Dias:2009ex}.
 
\subsection{Solution of the radial equation}\label{sec:radialeq}

In this subsection, we will find that the radial equation in NHEK-AdS can be solved exactly. This is a remarkable feature of perturbations in NHEK-AdS. 

The radial equation \eqref{NHgeometry_radial master eq} is an ODE with no singular points and three regular singular points at $\pm i$ and $\infty$. Therefore it can be transformed into the hypergeometric equation.
Introducing $\phi_{lm\omega}^{(s)}(z)=z^{\tilde{\alpha}}\left(1-z\right)^{\tilde{\beta}}F(z)\,$, with $z=\frac{1}{2}\left(1-i r\right)\,$, the radial equation \eqref{NHgeometry_radial master eq} exactly agrees with the hypergeometric equation
\begin{equation}\label{radialeq_hypergeometric eq}
z\left(1-z\right)\partial_z^2F(z)+\left[\tilde{c}-\left(\tilde{a}+\tilde{b}+1\right)z\right]\partial_zF(z)-\tilde{a}\tilde{b}F(z)=0\,,
\end{equation}
with the identifications
\begin{alignat}{3}
\tilde{\alpha}=&{\textstyle \frac{1}{2}}\left(\omega-i q\right)\,,\qquad&\tilde{\beta}=&{\textstyle \frac{1}{2}}\left(\omega+i q\right)\,,\qquad&\eta^2=&1+4\left(\mu^2-q^2\right)\,,\nonumber \\
\tilde{a}=&{\textstyle \frac{1}{2}}\left(1+\eta+2\omega\right)\,,\qquad& \tilde{b}=&{\textstyle \frac{1}{2}}\left(1-\eta+2\omega\right)\,,\qquad& \tilde{c}=&1+\omega-i q\,.
\end{alignat}
As \eqref{radialeq_hypergeometric eq} is symmetric under the interchange of $\tilde{a}$ and $\tilde{b}$, which differ merely by $\pm \eta\,$, we can, without loss of generality, demand $\eta\geq 0\,$, with 
\begin{equation}\label{radialeq_eta definition}
\eta\equiv \sqrt{1+4\left(\mu^2-q^2\right)}=\sqrt{1+4\tLambda_{lm}^{(s)}}\,.
\end{equation}
It follows from the discussion below \eqref{def:muq} that $\eta(-s)=\eta(s)$.
Given that none of the numbers $\tilde{c},\, (\tilde{c}-\tilde{a}-\tilde{b}),\, (\tilde{a}-\tilde{b})$ is equal to an integer \cite{abramowitz1964handbook} the most general solution of \eqref{NHgeometry_radial master eq}, in the neighbourhood of the regular  singular point $z=0$, reads
\begin{equation}\label{radialeq_full radial solution}
\phi_{lm\omega}^{(s)}(z)=A_0 \,z^{\tilde{\alpha}}\left(1-z\right)^{\tilde{\beta}}F(\tilde{a},\tilde{b},\tilde{c},z)+B_0\,z^{\tilde{\alpha}-\tilde{c}+1}\left(1-z\right)^{\tilde{\beta}}F(\tilde{a}-\tilde{c}+1,\tilde{b}-\tilde{c}+1,2-\tilde{c},z)\,.
\end{equation}
$A_0,B_0$ are constant amplitudes to be determined by boundary conditions. To render the function $\phi_{lm\omega}^{(s)}(z)$ single valued we choose the branch cuts $[-\infty,0]$ and $[1,+\infty]\,$, which corresponds to $|\arg(z)|<\pi$ and $|\arg(1-z)|<\pi\,$. Note that the above solution is regular for all finite values of $r$.

To further discuss the properties of the radial solution (and hence the physical properties of the perturbations) we first need to solve the angular equation to find its eingenvalues and thus determine $\eta$. We do this in the next subsection. Later, in Section \ref{sec:PhysInterp} we will return to \eqref{radialeq_full radial solution} and analyze its properties.

\subsection{Solution of the angular equation}\label{sec:AngSol}

To fully specify the radial solution \eqref{radialeq_full radial solution} we still need to determine the allowed values of the angular eigenvalues $\tLambda_{lm}$ defined in \eqref{def:muq}.
As will be shown in the next section, $\eta^2$ governs the behaviour of the solutions at infinity and its determination is therefore fundamental. 
We will therefore present our results for $\eta^2$; $\tLambda_{lm}$ can then be read from \eqref{radialeq_eta definition}. 

The value of $\tLambda_{lm}$, and thereby $\eta$, depends on four dimensionless  parameters: the quantum numbers $s$, $l$, $m$, which label the spin, the total angular momentum and its projection, and on the ratio $\nicefrac{r_+}{\ell}\,$. The latter quantity retains the memory of the horizon size in AdS units of the extreme Kerr-AdS whose near-horizon geometry is described by NHEK-AdS.

\begin{figure}[h]
\centering
\includegraphics[width=.48\textwidth]{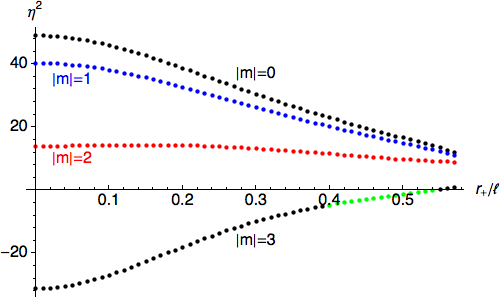}
\hspace{0.3cm}
\includegraphics[width=.48\textwidth]{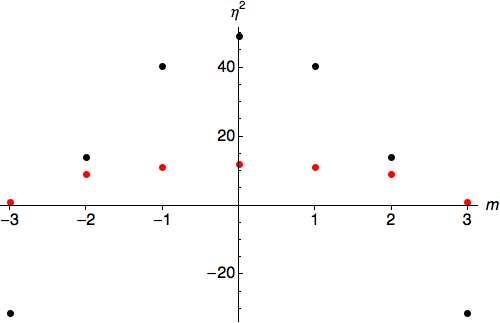}
\caption{$\eta^2$, defined in \eqref{radialeq_eta definition}, for $|s|=2$ and $l=3$. {\bf a)} $\eta^2$ vs $\nicefrac{r_+}{\ell}\,$ for $|m|=0,1,2,3$, and {\bf b)} $\eta^2$ vs $m$ for $\nicefrac{r_+}{\ell}=0\,$ (black points) and  $\nicefrac{r_+}{\ell}=0.55 \simeq \nicefrac{1}{\sqrt{3}}\,$ (red points).}\label{Fig:1}
\end{figure}

\begin{figure}[b]
\centering
\includegraphics[width=.48\textwidth]{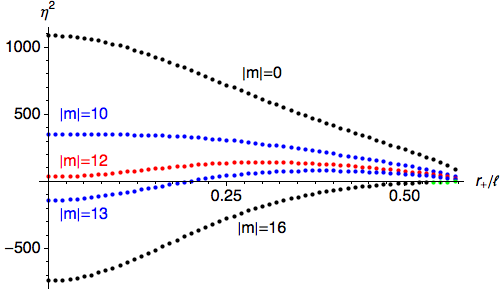}
\hspace{0.3cm}
\includegraphics[width=.48\textwidth]{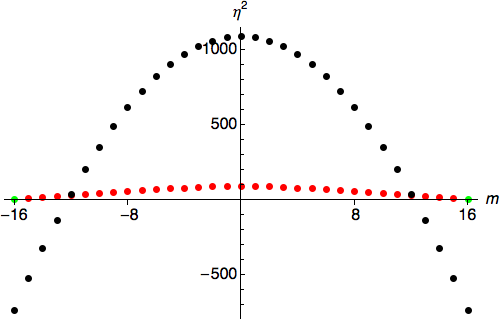}
\caption{$\eta^2$ for $|s|=2$ and $l=16$. {\bf a)} $\eta^2$ vs $\nicefrac{r_+}{\ell}\,$ for the representative $|m|$ cases (from top to bottom these are $|m|=0,10,12,13,16$), and {\bf b)} $\eta^2$ vs $m$ for $\nicefrac{r_+}{\ell}=0\,$ (black points) and  $\nicefrac{r_+}{\ell}=0.55 \simeq \nicefrac{1}{\sqrt{3}}\,$ (red points).}\label{Fig:2}
\end{figure}

 Recall also that $\nicefrac{a}{\ell}\,$ is fixed by the constraint \eqref{NHgeometry_extremality condition}.
As mentioned before, the quantum numbers $s$, $l$, $m$ are integers constrained to satisfy the regularity conditions  $-l\leq m\leq l$ and $l\geq|s|$, and the number of zeros of a specific eigenfunction $S^{(s)}_{lm}(\theta)$ is given by $l-\max\left\{|m|,|s|\right\}$. 

\begin{figure}[t]
\centering
\includegraphics[width=.47\textwidth]{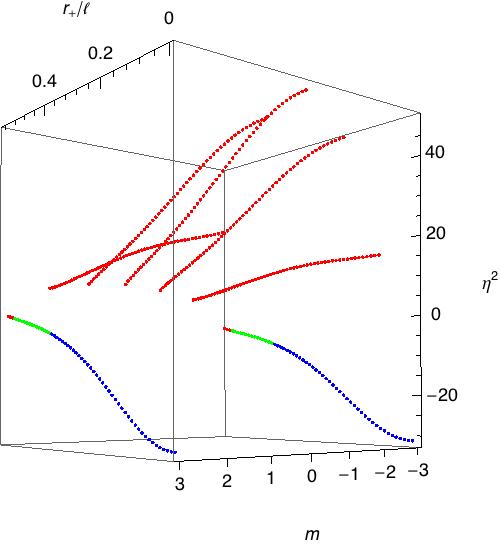}
\hspace{0.5cm}
\includegraphics[width=.47\textwidth]{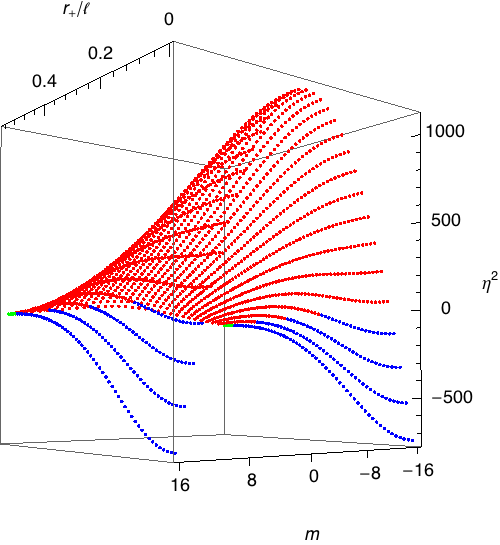}
\caption{$\eta^2$ as a function of $-l\leq m\leq l$ and  $\nicefrac{r_+}{\ell}\,$, for $l=3$ ({\it left}) and $l=16$ ({\it right}). The red points (curve segments) have $\eta^2>0$ while the blue points (curve segments) have $\eta^2<0$. In Section \ref{sec:PhysInterp} we will conclude that the red dots describe normal modes ($\eta\in\Reals$), while blue dots describe traveling waves ($\eta\in\Imaginaries$). The green dots correspond to modes on which we cannot impose outgoing boundary conditions.}\label{Fig:3}
\end{figure}
\begin{figure}[h]
\centering
\includegraphics[width=.6\textwidth]{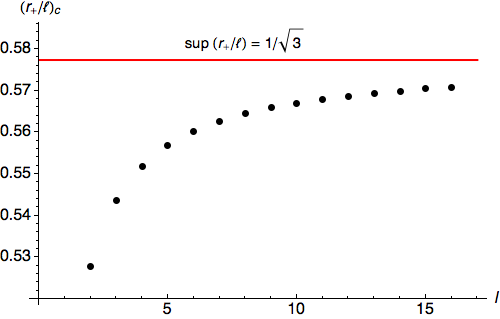}
\caption{Critical value $(\nicefrac{r_+}{\ell})_c\,$ for $l\leq16$. In Section \ref{sec:PhysInterp} we conclude that no traveling waves exist for $(r_+/\ell)_c<r_+/\ell<\sup\{r_+/\ell\}$.}\label{Fig:4}
\end{figure}

We use spectral methods to solve  the angular equation numerically. In contrast to finite difference and finite element methods, which use local trial functions, spectral methods use global trial functions. For analytical functions, spectral methods have exponential convergence properties. In a first step we employ the Frobenius method to map equation \eqref{NHgeometry_angular master eq}, which has regular singular points at $\theta=\pm\frac{\pi}{2}$, into a differential equation without singular points plus a set of boundary conditions at $\theta=\pm\frac{\pi}{2}$. We then  use a Chebyshev grid discretization. The problem boils down to a generalized eigenvalue equation for $\tLambda_{lm}$. This eigenvalue problem can readily be solved in {\it Mathematica}. As our focus lies on gravitational perturbations, all numerical calculations are performed for $|s|=2$ (recall that $\eta(-s)=\eta(s)$; in addition the eigenvalues are symmetric under the interchange $m\leftrightarrow-m$).

We have computed $\eta^2$ as a function of $m$ and of $\nicefrac{r_+}{\ell}\,$ for $2\leq l\leq30$ \footnote{The dimensionless horizon radius $\nicefrac{r_+}{\ell}\,$ is a continuous parameter; we choose a step size of $0.01$ in the presentation of our results.}.
Figures  \ref{Fig:1} and  \ref{Fig:2} are two representative examples of our results: Figure \ref{Fig:1} is for $l=3$ while Figure \ref{Fig:2} is for $l=16$.
In these figures the left panel gives $\eta^2$ as a function of  $\nicefrac{r_+}{\ell}\,$ for several fixed values of $m$. On the other hand, the right panel displays  $\eta^2$ as a function of  $m$ for two different radii, namely $\nicefrac{r_+}{\ell}=0\,$ (the flat limit) and  $\nicefrac{r_+}{\ell}=0.55 \lesssim \nicefrac{1}{\sqrt{3}}\,$(recall that, as discussed in \eqref{NHgeometry_extremality condition}, the metric is no longer well behaved for $r_+/\ell=1/\sqrt{3}$).
Finally, in the right panel of Figure \ref{Fig:3} we complete the information that is missing in Figure \ref{Fig:2} with a 3-dimensional plot that shows $\eta^2$ as a function of $-l \leq m\leq l$ and of $\nicefrac{r_+}{\ell}\,$, for $l=16$. For completeness, in the left panel we also show the equivalent plot for $l=3$. To understand the color code employed in these plots we anticipate some relevant information that will be discussed in detail in the next section. We will find that $\eta^2>0$ (red points in the 3-dimensional plots) corresponds to normal modes which decay at infinity, whereas $\eta^2<0$ (blue points in the 3-dimensional plots)  describes traveling waves which oscillate at infinity. Moreover, we will find a special sector of modes for which we cannot impose outgoing boundary conditions. These modes are identified by green dots in our plots.

These plots have some interesting properties. To start with, the points in the $\nicefrac{r_+}{\ell}= 0\,$ plane describe the asymptotically flat limit, $\ell\rightarrow\infty$. An important check of our numerical code, is that our calculations exactly reproduce the results presented in \cite{Dias:2009ex} for the NHEK geometry.
Note that in this case, $\eta^2$ can be positive (this happens for small values of $m$) or negative (for larger values of $m$).  A similar situation holds when $\nicefrac{r_+}{\ell}$ is non-vanishing but not too large (see further discussion below). Again, the sign of $\eta^2$ will play an important role in the physical interpretation of the perturbations done in the next section.

Next, fix $l$ and $m$ and follow the evolution of $\eta^2$ as $\nicefrac{r_+}{\ell}$ grows from zero to its upper bound $\sup\{r_+/\ell\}=1/\sqrt{3}$. In this path, if $\eta^2$ starts positive, it remains positive. This is the typical ``small" $m$ behaviour. In particular, $\eta^2$ is always positive  for $m=0$ modes that are relevant for the conjecture  \cite{Durkee:2010ea} discussed in the introduction.  On the other hand, if $\eta^2$ starts negative at $\nicefrac{r_+}{\ell}=0$, it does change sign at some  intermediate $\nicefrac{r_+}{\ell}$ before reaching $\nicefrac{r_+}{\ell}\to\nicefrac{1}{\sqrt{3}}\,$. Typically this happens for ``large" values of $m \lesssim l$ and as we approach the upper bound the modes with $|m|=l$ are the last to change sign.  Given an $l$ there is a critical dimensionless radius $\nicefrac{r_+}{\ell}=(\nicefrac{r_+}{\ell})_c <\nicefrac{1}{\sqrt{3}}$ above which $\eta^2$ is always positive for {\it any} $|m|\leq l$. (In the next section we will find that as a consequence there are no traveling waves for $(r_+/\ell)_c<r_+/\ell<1/\sqrt{3}$). This threshold is not universal, it depends on the quantum number $l$.
The evolution of this critical value $(\nicefrac{r_+}{\ell})_c $ as a function of the quantum number $l$ for $l\leq 16$ is illustrated in  Figure  \ref{Fig:4}. This value $(\nicefrac{r_+}{\ell})_c $ grows monotonically  approaching  $ \nicefrac{1}{\sqrt{3}}$ (where the metric is no longer well-behaved) as $l$ grows. For higher $l$, $(\nicefrac{r_+}{\ell})_c $ is closer to the singular value $ \nicefrac{1}{\sqrt{3}}$ and the numerical results become less accurate.

\section{Analysis of the solutions. Stability and  Kerr/CFT discussions}\label{sec:PhysInterp}

At this stage we have found the eigenvalues of the angular equation for perturbations in NHEK-AdS, which can be plugged in the exact radial solution \eqref{radialeq_full radial solution}. This radial solution depends on only one undetermined parameter, namely the frequency of the perturbation. It might be constrained by the asymptotic boundary conditions.  In Subsection \ref{sec:PIstability} we select a sector of boundary conditions and search (unsuccessfully) for linear unstable modes of NHEK-AdS.   
In particular, we do not find any axisymmetric instability, which is in agreement with a recent conjecture  \cite{Durkee:2010ea} (see introduction) relating the stability properties of the full geometry to those of its near-horizon geometry.  In Subsection \ref{kerrCFT}, we find that the asymptotic behaviour of the metric perturbations in NHEK-AdS violates the fall-off conditions imposed in the formulation of the Kerr/CFT correspondence (the only exception being the axisymmetric sector of perturbations).

\subsection{Boundary conditions. Search for unstable modes of NHEK-AdS\label{sec:PIstability}}

NHEK-AdS has timelike asymptotic boundaries at $r=\pm\infty$ and, having the exact analytical solution \eqref{radialeq_full radial solution} for the radial perturbation $\phi_{lm\omega}^{(s)}$,  we can find its asymptotic behavior.
We use standard properties of the hypergeometric functions \cite{abramowitz1964handbook} to map the regular singular point $z=0$ onto the regular singular point $z=1$. We further employ the series expansion of the hypergeometric function and of the exponential function. The desired asymptotic behaviour, to next-to-leading order, is
\begin{equation}\label{radialeq_behaviour at infinity}
\begin{split}
\underset{r\rightarrow\pm\infty}{\lim}\phi^{(s)}_{lm\omega}(r)\sim&\,2^{\frac{1+\eta}{2}}\Gamma(\tilde{b}-\tilde{a})C^{\pm}e^{\pm i\pi(\tilde{\beta}-\tilde{\alpha}-\tilde{a})}e^{-\frac{1+\eta}{2}\ln|r|}e^{-\frac{2q\omega}{1+\eta}\frac{1}{r}}\\
&+2^{\frac{1-\eta}{2}}\Gamma(\tilde{a}-\tilde{b})D^{\pm}e^{\pm i\pi(\tilde{\beta}-\tilde{\alpha}-\tilde{b})}e^{-\frac{1-\eta}{2}\ln|r|}e^{-\frac{2q\omega}{1-\eta}\frac{1}{r}}\,,
\end{split}
\end{equation}
where
\begin{align}\label{defCD}
C^{\pm}=&A_0\,\frac{\Gamma(\tilde{c})}{\Gamma(\tilde{b})\Gamma(\tilde{c}-\tilde{a})}-B_0\,e^{\pm i\pi \tilde{c}}\frac{\Gamma(2-\tilde{c})}{\Gamma(\tilde{b}-\tilde{c}+1)\Gamma(1-\tilde{a})}\,,\nonumber \\
D^{\pm}=&A_0\,\frac{\Gamma(\tilde{c})}{\Gamma(\tilde{a})\Gamma(\tilde{c}-\tilde{b})}-B_0\,e^{\pm i\pi \tilde{c}}\frac{\Gamma(2-\tilde{c})}{\Gamma(\tilde{a}-\tilde{c}+1)\Gamma(1-\tilde{b})}\,.
\end{align}
It follows from  \eqref{radialeq_eta definition} and the analysis of Section \ref{sec:AngSol} that, depending on the value of $\tLambda_{lm}^{(s)\,}$, $\eta$ can be either real or imaginary. The boundary condition discussion now depends on each of these two families of modes we look at. 

\subsubsection{Normal modes}

For $\eta\in\Reals$ we demand the solution to be normalizable (i.e. that the mode has finite energy), which means $D^\pm$ must vanish in \eqref{radialeq_behaviour at infinity}. This gives a pair of conditions for the amplitudes $A_0,\,B_0$ in the radial solution \eqref{radialeq_full radial solution}. 
Non-trivial solutions exist when the determinant of this system of equations vanishes, i.e.\footnote{To get the quantization conditions of this section, we use the Gamma function property  $\Gamma(z)\Gamma(1-z)=\pi/\sin(\pi z)$.}
\begin{equation}\label{radialeq_normal modes quantization condition}
\det=\frac{(1-\tilde{c})}{\Gamma(\tilde{a})\Gamma(\tilde{c}-\tilde{b})\Gamma(\tilde{a}-\tilde{c}+1)\Gamma(1-\tilde{b})}=0\,.
\end{equation}
Neither $(\tilde{c}-\tilde{b})$ nor $(\tilde{a}-\tilde{c}+1)$ depends on $\omega$, so this condition can be obeyed only if we use the property $\Gamma(-n)=\infty,\;n\in\Naturals_0\,$ to get the following frequency quantization,
\begin{equation}
\begin{array}{rlcl}
\tilde{a}=&\!\!\!\!-n\;&\Rightarrow\;&\omega=-\left(n+{\textstyle \frac{1}{2}+\frac{\eta}{2}}\right), \; n\in\Naturals_0\,; \quad B_0=0\; \\ \noalign{\smallskip}
(1-\tilde{b})=&\!\!\!\!-n\;&\Rightarrow\;&\omega=n+{\textstyle \frac{1}{2}+\frac{\eta}{2}}\,,  \; n\in\Naturals_0\,; \quad A_0=0
\end{array}
 \;\Biggr\}\rightarrow\;\;\omega=\pm\biggl(n+\frac{1}{2}+\frac{\eta}{2}\biggr)\,,\; n\in\Naturals_0 \,,
\end{equation}
where the last expression compiles the normal mode spectrum that arises from the two possible cases.
When $\ell\to \infty$ this spectrum agrees with the normal modes results of \cite{Dias:2009ex} and, in agreement with the discussion above,  it is precisely the spectrum of normal modes found for a massive charged  scalar in $AdS_2$ with a homogeneous electric field in Ref. \cite{Strominger:1998yg}.

In the above analysis we must distinguish the positive and negative frequency cases because the Teukolsky equations for $s\neq0$ are not invariant under complex conjugation. Therefore negative frequency solutions cannot simply be obtained from positive ones. They have to be considered separately and the two signs correspond to different helicities of the field \cite{Dias:2009ex}.

Naively $\tilde{c}=1$ would also satisfy the quantization condition \eqref{radialeq_normal modes quantization condition}. Yet, as mentioned above, the function \eqref{radialeq_full radial solution} no longer solves the radial equation if $c$ is an integer. When repeating the analysis with the appropriate regular solution \cite{abramowitz1964handbook}, we found that the special case $\tilde{c}=1$ has no physical relevance.

\subsubsection{Traveling waves}

For $\eta=i\teta\in\Imaginaries$, the solution describes traveling waves. Indeed, in this case the radial function oscillates at infinity and thus we can have incoming or outgoing waves. As discussed in association  with Figure \ref{Fig:3}, for a given $l$ there are no traveling waves when $\nicefrac{r_+}{\ell}>(\nicefrac{r_+}{\ell})_c$, but in the complementary regime (which includes the flat limit case $\ell\to \infty$) they do exist. 

 We are interested in studying the stability of the NHEK-AdS geometry against small perturbations but, in general, not in scattering experiments. Therefore, at each of the two asymptotic boundaries of our spacetime, we will require that we have only outgoing waves. There exist two different notions of  ``outgoing" depending on whether we discuss the phase or the group velocity, and these need not have the same sign. The latter governs the transmission speed of information and thus it is the physically relevant velocity. On the other hand, the phase velocity dictates the direction of the energy flux (i.e. for $\omega>0$ the energy flux has the same sign as the phase velocity). Since our modes have time-dependence of the form $e^{-i\,\omega\, t}$, a solution with positive frequency imaginary part has an amplitude that grows in time $-$ it describes an instability $-$ while a solution with negative imaginary part for the frequency is damped in time $-$ it is a quasinormal mode.

To determine the group and phase velocity, revisit equation \eqref{radialeq_behaviour at infinity} and define
\begin{equation}\label{defS}
\mathcal{S}^{\nicefrac{C}{D}}=i\left[\mp\frac{\teta}{2}\ln|r|+\frac{2\omega\left(\pm q_0\,\teta+s\right)}{1+\teta^2}\frac{1}{r}\right], \qquad  q_0\equiv{\rm Re}(q)=\frac{2a\,m\,r_+\Xi}{V\left(r_+^2+a^2\right)}.
\end{equation}
Here (and in the expressions below for $v_{ph}^{\nicefrac{C}{D}}$ and $v_{gr}^{\nicefrac{C}{D}}$) the superscript $\nicefrac{C}{D}$ refers to the upper/lower sign in the RHS of the respective expression. Moreover, the subscripts in $C^\pm$ and $D^\pm\,$ defined in \eqref{radialeq_behaviour at infinity} (and used in Table \ref{Table:CDsigns}) are associated with $r\to \pm \infty$. With the definition \eqref{defS}, $e^{\,\mathcal{S}^{\nicefrac{C}{D}}}$ describes the radial contribution to the wave propagation in the context of a WKB (Wentzel-Krames-Brillouin) approximation analysis. Introducing the WKB effective wave number $k^{\nicefrac{C}{D}}(r)=-i\,\partial_r\mathcal{S}^{\nicefrac{C}{D}}\,$, the phase and group velocity are then, respectively, given by
\begin{equation}
v_{ph}^{\nicefrac{C}{D}}=\frac{\omega}{k^{\nicefrac{C}{D}}}\sim\mp\frac{2\omega}{\teta}\,r\,,\qquad v_{gr}^{\nicefrac{C}{D}}=\left(\frac{dk^{\nicefrac{C}{D}}}{d\omega}\right)^{-1}\!\!\!\sim\mp\,\frac{1}{2}\,\frac{\left(1+\teta^2\right)}{\left(q_0\,\teta\pm s\right)}\,r^2\,.
\end{equation}
At $r = \pm \infty$, depending on which subset of the  amplitudes $\{C_\pm,D_\pm \}$ we set to zero,  we can have the combinations for the sign of the phase and group velocities displayed in Table \ref{Table:CDsigns}.
Again, we will consider only cases describing outgoing boundary conditions at both boundaries of NHEK-AdS. Modes described by the two last rows of Table \ref{Table:CDsigns} cannot obey such boundary conditions. These are the modes identified with the green color in the eigenvalue plots shown in Figures \ref{Fig:1}-\ref{Fig:3} of Section  \ref{sec:AngSol}.
\begin{table}[h]
\centering
\begin{tabular}{|c|c|c|c|c|c|}
\cline{3-6}
\multicolumn{2}{c|}{}&$C^+$&$D^+$&$C^-$&$D^-$ \\ \hline
\multirow{2}{*}{$v_{ph}$}&${\rm Re}(\omega)>0$& $-$ & $+$ & $+$ & $-$ \\ \cline{2-6}
&${\rm Re}(\omega)<0$& $+$ & $-$ & $-$ & $+$ \\ \hline
\multirow{2}{*}{$v_{gr}$}&$q_0\teta\mp s>0$& $-$ & $+$ & $-$ & $+$ \\ \cline{2-6}
&$q_0\teta\mp s<0$& $+$ & $-$ & $+$ & $-$ \\ \cline{2-6}
&$q_0\teta+ s>0\,\text{, }q_0\teta- s<0 $& $-$ & $-$ & $-$ & $-$ \\ \cline{2-6}
&$q_0\teta+ s<0\,\text{, }q_0\teta- s>0 $& $+$ & $+$ & $+$ & $+$ \\ \hline
\end{tabular}
\caption{Signs of the amplitudes $C^\pm$ and $D^\pm\,$ introduced in \eqref{radialeq_behaviour at infinity}. They are needed to determined the signs of the phase and group velocity (see discussion in the text).}\label{Table:CDsigns}
\end{table}\\

Consider first the case where we look into boundary conditions where only outgoing phase velocity is allowed at both boundaries $r\to \pm \infty$. Bardeen and Horowitz identified this type of boundary condition as a case where there is room for a possible instability $-$ the ergoregion instability \cite{Friedman:1978} $-$ in near-horizon geometries since these are horizonless but have an ergoregion. In the flat case, \cite{Dias:2009ex} found however that no such instability is present in NHEK. Here we will conclude that a similar result holds for NHEK-AdS.  
We have to initially distinguish the positivity of the real part of the frequency, ${\rm Re}(\omega)$. For ${\rm Re}(\omega)>0$, from Table \ref{Table:CDsigns} we conclude that outgoing phase velocity at $r\to \pm \infty$ requires $C^{\pm}=0$. For ${\rm Re}(\omega)<0$ we have instead to set $D^{\pm}=0\,$.
The requirement $C^{\pm}=0$ boils down to the condition
\begin{equation}
\frac{(1-\tilde{c})}{\Gamma(\tilde{b})\Gamma(\tilde{c}-\tilde{a})\Gamma(\tilde{b}-\tilde{c}+1)\Gamma(1-\tilde{a})}=0\,,
\end{equation}
which is identical to \eqref{radialeq_normal modes quantization condition}, up to the interchange $\tilde{a}\leftrightarrow \tilde{b}$. So the quantization proceeds analogously to the treatment of the normal modes. Note that $\tilde{b}=-n$ is in conflict with ${\rm Re}(\omega)>0\,$, and therefore the only solution is $(1-\tilde{a})=-n\;\Rightarrow\;\omega=n+\frac{1}{2}-i\,\frac{\teta}{2}\,,\: n\in\Naturals_0$. For ${\rm Re}(\omega)<0\,$, the requirement $D^\pm=0$ can again be treated analogously to the case of the normal modes. We find that a possible solution $(1-\tilde{b})=-n$ for non-negative $n$ is not compatible with ${\rm Re}(\omega)<0$, and thus the only solution is 
$\tilde{a}=-n\;\Rightarrow\;\omega=-\left(n+\frac{1}{2}\right)-i\,\frac{\teta}{2}\,,\: n\in\Naturals_0$.
We can summarize the two frequency quantizations in the single result, 
\begin{equation}\label{radialeq_phase velocity quantization}
\omega=n+\frac{1}{2}-i\,\frac{\teta}{2}\,,\quad n\in\Integers\,.
\end{equation}
These are quasinormal modes of NHEK-AdS since the imaginary part of the frequency spectrum is negative.
To interpret this result recall the argument of Bardeen-Horowitz for the possible existence of an instability in this sector of perturbations. We required only outgoing phase so our perturbations (for positive frequency modes) have necessarily outgoing energy flux at infinity. But NHEK-AdS has an ergoregion 
where negative energy states are allowed, and thus where the Penrose process and superradiant emission can occur. So if we start with some localized initial data with negative energy and if a perturbation removes energy from such a system, the energy at the ergoregion core could grow negatively large and lead to an instability \cite{friedman}.    
However, we have found that outgoing phase always leads to stable quasinormal modes rather than an instability, like in the flat limit of our analysis.
The reason for the absence of the instability was identified in the NHEK case in \cite{Dias:2009ex}, and also holds when the cosmological constant is present.
Take the ${\rm Re}(\omega)>0,\,  q_0\,\teta\mp s>0$ case for concreteness (the description for the other cases is similar). Imposing $C^{+}=0$ means that at $r\to \infty$ both the phase and group velocities have the same sign. On the other hand, the condition $C^{-}=0$ means that at $r\to -\infty$ we have outgoing phase but the group velocity is ingoing: we have energy flux leaving the spacetime through this boundary but this corresponds to the physical propagation of an a incoming wave. Thus, we have a very fine-tuned (and in this sense unphysical) experiment: we prepare our initial data to be such that an initial wavepacket (at finite $r$ in the the bulk of the geometry) does not propagate to $r= -\infty$ by sending in an appropriate (finely tuned) wavepacket from $r= -\infty$ to scatter
with it in such a way as to produce only a wavepacket propagating to
$r = +\infty$.\footnote{The analogous situation in a Kerr black hole would be boundary conditions where one manipulates the initial data to be such that no waves
cross the future horizon by sending in appropriate and finely tuned waves from the
past horizon.} This fine-tuning is probably the reason that we do
not see an instability in NHEK \cite{Dias:2009ex} or in NHEK-AdS.

Consider now the physical case where we impose outgoing group velocity boundary conditions at both boundaries.
From Table  \ref{Table:CDsigns}, these boundary conditions require either $C^+=D^-=0$, if  $q_0\,\teta\mp s>0$, or $C^-=D^+=0$,  if  $q_0\,\teta\mp s<0$ (note that the cases described in the two last rows of  Table  \ref{Table:CDsigns}  can never describe a system with outgoing group velocity at {\it both} boundaries).  This pair of conditions translates, respectively, into the  quantization conditions 
\begin{align}
\sin(\pi \tilde{b})\sin\bigl[\pi(\tilde{c}-\tilde{a})\bigr]e^{-i\pi \tilde{c}}=&\sin(\pi \tilde{a})\sin\bigl[\pi(\tilde{c}-\tilde{b})\bigr]e^{i\pi \tilde{c}}\,,\nonumber \\
\sin(\pi \tilde{a})\sin\bigl[\pi(\tilde{c}-\tilde{b})\bigr]e^{-i\pi \tilde{c}}=&\sin(\pi \tilde{b})\sin\bigl[\pi(\tilde{c}-\tilde{a})\bigr]e^{i\pi \tilde{c}}\,,
\end{align}
which can be solved with the help of {\it Mathematica}. The solutions of these two cases combine to give the single frequency quantization
\begin{equation}
\omega=n+\frac{1}{2}-\frac{i}{2\pi}\ln\biggl[\frac{\cosh\left[\pi\left(\teta/2+|q_0|\right)\right]}{\cosh\left[\pi\left(\teta/2-|q_0|\right)\right]}\biggr],\quad n\in\Integers\,,
\end{equation}
where we have restricted our analysis to the most relevant spins $|s|=0,2\,$.
As ${\rm Im}(\omega)<0$ these solutions are damped, i.e. these are quasinormal modes of NHEK-AdS.

To sum up this Subsection \ref{sec:PIstability}, in a linear mode search for instabilities in NHEK-AdS that have outgoing boundary conditions, we do not find any sign of unstable modes. (However, we cannot rule out the possibility that a different set of boundary conditions might lead to an instability). 
This applies both to normal waves and traveling modes and both to non-axisymmetric and axisymmetric modes. As discussed in the Introduction, the fact that we do not find an axisymmetric instability in NHEK-AdS is in agreement with the conjecture proposed in  \cite{Durkee:2010ea}, and here verified for the Kerr-AdS system. Recall that the modes relevant for this conjecture are the normal modes ($\eta^2>0$) with $m=0$. 

\subsection{Hertz map for metric perturbations. Implications for the Kerr/CFT correspondence\label{kerrCFT}}

Many physically interesting quantities can be directly computed from the gauge invariant Weyl scalars of the Newman-Penrose formalism. Yet, for some problems, it is essential to know  the linear perturbation $h_{\mu\nu}\,$ of the metric itself. The Hertz map, $h_{\mu\nu}=h_{\mu\nu}(\Psi_H)$, reconstructs the perturbations of the metric tensor (or of the electromagnetic vector potential) from the associated scalar Hertz potentials $\Psi_H$ (in a given gauge). These are themselves closely related to the Weyl scalar perturbations discussed in the previous sections.
The Hertz map construction studies have been pioneered by Cohen, Kegeles and Chrzanowski \cite{Cohen19755,Kegeles:1979an,Chrzanowski:1975wv} and were further explored by Stewart \cite{Stewart:1978tm}. Wald \cite{Wald:1978vm} revisited the problem and provided  an elegant and straightforward proof of the relation between the perturbation equations for the Weyl scalars  and the corresponding Hertz potentials. A brief but complete  review of the subject can be found in an appendix of \cite{Dias:2009ex}.
Here we apply this Hertz map to our problem.

For vacuum type D spacetimes, the Hertz potential itself satisfies a master equation, which is also the basis for its definition. More specifically, the Hertz potentials  obey the second order differential equations ($s^\pm=\pm\frac{1}{2},\,\pm\frac{3}{2},\,\pm1,\,\pm2$)
\begin{subequations} \label{HertzMaster}
\begin{align}
&{\bigl \{} \left[\Delta-\left(2s^-\!+1\right)\gamma-\bgamma+\bmu\right]\left(D -2s^-\epsilon-\left(2s^-\!+1\right)\rho\right)-\left[\bdelta-\btau+\bbeta-\left(2s^-\!+1\right)\alpha\right] \nonumber \\
&\:\times\left(\delta-\left(2s^-\!+1\right)\tau-2s^-\beta\right)+{\textstyle \frac{1}{3}}s^-\bigl(s^-\!+{\textstyle \frac{1}{2}}\bigr)\left(s^-\!+1\right)\left(2s^-\!+7\right)\Psi_2 \,{\bigr \}}\, \psi_{H}^{(s^-)}=0 \,,\\
&{\bigl \{}  \left[D-\left(2s^+\!-1\right)\epsilon+\bepsilon-\brho\right]\left(\Delta-\left(2s^+\!-1\right)\mu-2s^+\gamma\right)-\left[\delta+\bpi-\balpha-\left(2s^+\!-1\right)\beta\right]  \nonumber \\
&\:\times\left(\bdelta-\left(2s^+\!-1\right)\pi-2s^+\alpha\right)+{\textstyle \frac{1}{3}}s^+\bigl(s^+\!-{\textstyle \frac{1}{2}}\bigr)\left(s^+\!-1\right)\left(2s^+\!-7\right)\Psi_2 \,{\bigr \}}\,\psi_{H}^{(s^+)}=0 \,.
\end{align}
\end{subequations}
In the special case of the NHEK-AdS (or the Kerr-AdS) geometry, the Hertz potential obeys the same master equation as its conjugated Teukolsky field but with spin sign traded.\footnote{A comparison between the conjugate relations \eqref{NHgeometry_redef scalars}-\eqref{ansatzWeyl} and \eqref{HertzAnsatz} clarifies this statement.}
That is, if we replace
\begin{equation}
\psi_{H}^{(s)}= \left\{
\begin{array}{ll}
e^{-i\omega t}e^{i m\phi} \left(1+r^2\right)^{-s/2}\Phi^{(s)}_{lm\omega}(r) S_{lm}^{(s)}(\theta)\,, & \qquad s\leq 0 \,,\label{HertzAnsatz} \\
e^{-i\omega t}e^{i m\phi}  \left(1+r^2\right)^{-s/2}\Phi^{(s)}_{lm\omega}(r)  S_{lm}^{(s)}(\theta) \left(
-\Psi_{2}\right)^{-\frac{2s}{3}} \,, & \qquad s\geq 0 \,,
\end{array}
\right.
\end{equation}
into \eqref{HertzMaster} we find that $\Phi^{(s)}_{lm\omega}(r)$ and $S_{lm}^{(s)}(\theta)$ are exactly the solutions of the radial equation \eqref{NHgeometry_radial master eq} and of the angular equation \eqref{NHgeometry_angular master eq}, respectively. 

Onwards we are interested only in spin $s=\pm 2$ perturbations and thus we restrict our analysis to the gravitational Hertz map.
The Hertz potentials $\psi_{H}^{(-2)}$ and $\psi_{H}^{(2)}$ contain the same physical information. Through the Hertz map they generate the metric perturbations in two different gauges, namely the ingoing (IRG) and the outgoing (ORG) radiation gauge, defined by
\begin{equation}
\mathrm{IRG}:\;\NPl^{\mu}h_{\mu\nu}=0,\;g^{\mu\nu}h_{\mu\nu}=0\,,\qquad\mathrm{ORG}:\;\NPn^{\mu}h_{\mu\nu}=0,\;g^{\mu\nu}h_{\mu\nu}=0\,.
\end{equation}
For a detailed discussion of the definition and existence of radiation gauges in Petrov type II and D spacetimes see \cite{Price:2006ke}. The resulting linear perturbations of the metric are given by\footnote{\label{footMap}Note that \eqref{Hertzmap_metric perturbations ORG}, whose explicit derivation can be found in an Appendix of \cite{Dias:2009ex}, corrects some typos in the map first presented in \cite{Chrzanowski:1975wv}.}
\begin{eqnarray}\label{Hertzmap_metric perturbations IRG}
&&h_{\mu\nu}^{\mathrm{IRG}}=\Bigl\{\NPl_{(\mu}\NPm_{\nu)}\bigl[\left(D+3\epsilon+\bepsilon-\rho+\brho\right)\left(\delta+4\beta+3\tau\right)+\left(\delta+3\beta-\balpha-\tau-\bpi\right)\left(D+4\epsilon+3\rho\right)\bigr]
\nonumber \\
&&\hspace{1cm}-\NPl_{\mu}\NPl_{\nu}\left(\delta+3\beta+\balpha-\tau\right)\left(\delta+4\beta+3\tau\right)-\NPm_{\mu}\NPm_{\nu}\left(D+3\epsilon-\bepsilon-\rho\right)\left(D+4\epsilon+3\rho\right)\Bigr\}\psi_H^{(-2)} 
\nonumber \\
&&\hspace{1cm}+\text{c.c.}\,,\\
&&h_{\mu\nu}^{\mathrm{ORG}}=\Bigl\{\NPn_{(\nu}\NPbm_{\mu)}\bigl[\left(\bdelta+\bbeta-3\alpha+\btau+\pi\right)\left(\Delta-4\gamma-3\mu\right)+\left(\Delta-3\gamma-\bgamma+\mu-\bmu\right)\left(\bdelta-4\alpha-3\pi\right)\bigr]
\nonumber \\
&&\hspace{1cm}-\NPn_{\mu}\NPn_{\nu}\left(\bdelta-\bbeta-3\alpha+\pi\right)\left(\bdelta-4\alpha-3\pi\right)-\NPbm_{\mu}\NPbm_{\nu}\left(\Delta-3\gamma+\bgamma+\mu\right)\left(\Delta-4\gamma-3\mu\right)\Bigr\}\psi_H^{(2)} \nonumber \\
&&\hspace{1cm}+\text{c.c.}\,.
\label{Hertzmap_metric perturbations ORG}
\end{eqnarray}
We have explicitly checked that \eqref{Hertzmap_metric perturbations IRG} and \eqref{Hertzmap_metric perturbations ORG} satisfy the linearized Einstein equations for traceless perturbations \cite{Dias:2009ex} (see also footnote \ref{footMap}).

In the context of the Kerr/CFT proposal, we are now interested in the asymptotic fall-off of the metric perturbation in NHEK-AdS. This can be obtained using  the Hertz map \eqref{Hertzmap_metric perturbations IRG} and \eqref{Hertzmap_metric perturbations ORG}, and the asymptotic expansion \eqref{radialeq_behaviour at infinity} for the radial function $\Phi^{(s)}_{lm\omega}$. Here one has to be cautious with a possible regularity issue: the basis vector fields  $\NPl$ and $\NPn$ are globally well-defined, but the vector field $\NPm$ is singular at $\theta=0,\pi$. However, this is harmless since the  angular dependence of the Hertz potential has a sufficiently high power of $\sin \theta$ to
ensure smoothness of $h_{\mu\nu}$ at $\theta=0,\pi$.
We find that the asymptotic result is independent of whether we work in the ingoing or outgoing radiation gauge. The explicit asymptotic behaviour of the metric perturbation is 
\begin{equation}\label{Hertzmap_Kerr-AdS metric perturbations}
h_{\mu\nu}^{GR}\sim r^{\frac{3}{2}\pm\frac{\eta}{2}}
\left (
\begin{array}{cccc}
\noalign{\vspace{5pt}}
\mathcal{O}\left(1\right)&\mathcal{O}\left(\frac{1}{r^2}\right)&\mathcal{O}\left(\frac{1}{r}\right)&\mathcal{O}\left(\frac{1}{r}\right)\vspace{0.2cm}\\
& \mathcal{O}\left(\frac{1}{r^4}\right)&\mathcal{O}\left(\frac{1}{r^3}\right)&\mathcal{O}\left(\frac{1}{r^3}\right)\vspace{0.2cm}\\
& &\mathcal{O}\left(\frac{1}{r^2}\right)&\mathcal{O}\left(\frac{1}{r^2}\right)\vspace{0.2cm}\\
& & &\mathcal{O}\left(\frac{1}{r^2}\right)
\vspace{5pt}
\end{array}
\right)\,,
\end{equation}
where the  rows and columns follow the sequence $\{t,r,\theta,\phi\}$. At this point we have not yet imposed  any boundary conditions, and recall that $\eta$ is the quantity related to the AdS spheroidal harmonic eigenvalue defined in \eqref{radialeq_eta definition}. 

We now want to compare the above  asymptotic behaviour of the metric
perturbations with the Kerr/CFT  fall-off conditions. Contrary to \eqref{Hertzmap_Kerr-AdS metric perturbations}, where $\eta$ in the power of $r$ depends on the cosmological background, the  Kerr/CFT  fall-off conditions are the same for NHEK and NHEK-AdS and given by \cite{Guica:2008mu,Lu:2008jk}
\begin{equation}\label{Hertzmap_fall-off conditions}
h_{\mu\nu}^{Kerr/CFT}\sim
\left (
\begin{array}{cccc}
\noalign{\vspace{5pt}}
\mathcal{O}\left(r^2\right)&\mathcal{O}\left(\frac{1}{r^2}\right)&\mathcal{O}\left(\frac{1}{r}\right)&\mathcal{O}\left(1\right)\vspace{0.2cm}\\
& \mathcal{O}\left(\frac{1}{r^3}\right)&\mathcal{O}\left(\frac{1}{r^2}\right)&\mathcal{O}\left(\frac{1}{r}\right)\vspace{0.2cm}\\
& &\mathcal{O}\left(\frac{1}{r}\right)&\mathcal{O}\left(\frac{1}{r}\right)\vspace{0.2cm}\\
& & &\mathcal{O}\left(1\right)
\vspace{5pt}
\end{array}
\right)\,.
\end{equation}
The fundamental question is whether  these fall-off conditions are compatible with the  decays permitted by the linearized Einstein equation. Clearly, the biggest conflict between these two decays happens in the $tr$ and
$t\theta$ components. To have compatibility between \eqref{Hertzmap_Kerr-AdS metric perturbations}
and \eqref{Hertzmap_fall-off conditions}  in these components, $\eta$ must be real, so traveling waves are automatically excluded from the system if the Kerr/CFT fall-off is imposed. Real $\eta$ means that we use normalizable boundary conditions  (i.e. the
lower sign choice in \eqref{Hertzmap_Kerr-AdS metric perturbations}) and we need $\eta \ge 3$, if all the normal modes are to respect the Kerr/CFT fall-off.
However, in Section \ref{sec:AngSol} we found that there are many normal modes with $\eta<3$; e.g. we found the value of $\eta=0.03240$ for $l=|m|=2$ at $\nicefrac{r_+}{\ell}=0.5279$, and  $\eta=0.4242$ for $l=|m|=3$ and $r_+/\ell=0.55$.
The conclusion of this analysis is that the Kerr/CFT fall-off conditions exclude all traveling waves and some normal modes from the spectrum of allowed perturbations. 

As observed in \cite{Dias:2009ex}, we could argue that a  gauge transformation could map a
mode violating the fall-off conditions onto one that satisfies these
conditions. However, this seems unlikely, especially for traveling waves.
We could also restrict our choice of initial data to a set of linear normal modes that satisfies 
the fall-off conditions but at the non-linear level their interaction will most likely excite traveling modes 
($\eta^2<0$)  that will violate the Kerr/CFT fall-off conditions. Considering a further possibility, a sum of the ingoing and outgoing radiation gauge perturbations (plus a diffeomorphism) does not obey the Kerr/CFT fall-off conditions.

In the NHEK geometry, Ref.  \cite{Dias:2009ex} observed that the only modes that could evade this conclusion
are the axisymmetric gravitational modes ($m=0$, $l\geq 2$) which have $\eta =
2l+1>3$. So they do obey the  Kerr/CFT fall-off conditions and they form  a consistent truncation of the full set of modes since linearized axisymmetric modes do not excite non-axisymmetric modes at next order in perturbation theory. We find that the same conclusion holds when $\ell$ is finite, i.e. in the $m=0$ sector, we always have $\eta>3$ for $0 \leq \nicefrac{r_+}{\ell}<  \nicefrac{1}{\sqrt{3}}\,$ (at least for the cases $2\leq l\leq30$ we verified). This is illustrated for the $l=3$ and $l=16$ cases in Figures \ref{Fig:1}-\ref{Fig:3}: for $m=0$ 
one has $\eta =2l+1>3$ for $\nicefrac{r_+}{\ell}=0$ and then it decreases as $\nicefrac{r_+}{\ell}$ grows. But in its way up to  $\nicefrac{r_+}{\ell}\to\nicefrac{1}{\sqrt{3}}$, $\eta$ stays well above the critical value of $3$.

\section*{Acknowledgments}
It is a pleasure to thank Malcolm J. Perry for helpful discussions. OD thanks the Yukawa Institute for Theoretical Physics (YITP) at Kyoto University, where part of this work was completed during the YITP-T-11-08 programme  ``Recent advances in numerical and analytical methods for black hole dynamics", and the  participants of the workshops ``Numerical Relativity and High Energy Physics", Madeira (Portugal) and ``Recent Advances in Gravity", Durham (UK) for discussions.  JS acknowledge support from NSF Grant No. PHY12-05500. MS acknowledges financial support from the British EPSRC Research Council, the German Academic Exchange Service, and the Cambridge European Trust. 
\begin{appendix}
\section{Newman-Penrose formalism and Teukolsky equations}\label{app:NP & Teukolsky}
In this appendix we will provide a short summary of the Newman-Penrose formalism and the Teukolsky perturbation equations including all formulae which are needed to derive our results in the main part of the paper. Teukolsky's original work only explicitly considers vacuum spacetimes, but his formalism is valid for any Petrov type-D background (like Kerr-AdS and NHEK-AdS).
\subsection{Newman-Penrose formalism}\label{sec:NPnutshell}

The Newman-Penrose (NP) formalism is suited to study dynamics in spacetimes that have at
least one preferred null direction, e.g. type D backgrounds like Kerr-AdS and near-horizon
Kerr-AdS.

The formalism requires a tetrad basis which consists of a pair of real null vectors $e_1=\NPl$, $e_2=\NPn$ and a pair of complex conjugate null vectors $e_3=\NPm$, $e_4=\NPbm\,$. The vectors obey the orthogonality relations $\NPl \cdot \NPm\! =\!\NPl\cdot \NPbm \!=\!\NPn \cdot \NPm\!=\!\NPn \cdot \NPbm\! =\!0$ and are normalized according to $\NPl \cdot \NPn\!=\!-1,\, \NPm \cdot \NPbm\!=\!1\,$.\footnote{The sign of both the normalization relations and the definition of all complex scalars in the Newman-Penrose formalism is related to the signature of the metric. The equations of the formalism, however, are independent of the metric signature. The definitions presented in this appendix are tied to the signature $(-,+,+,+)$.}
The Newman-Penrose formalism uses the tetrad basis to define directional derivative operators $D=\NPl^{\mu}\nabla_{\mu},\,\Delta=\NPn^{\mu}\nabla_{\mu},\,\delta=\NPm^{\mu}\nabla_{\mu},\,\bdelta=\NPbm^{\mu}\nabla_{\mu}\,$. We will label spacetime indices with Greek letters and tetrad indices with Latin letters. The central parameters of the formalism are three sets of complex scalars, defined as linear combinations of components of the Weyl tensor, the Ricci tensor and the spin connection $\gamma_{cab}=e_b^{\phantom{b}\mu}e_c^{\phantom{c}\nu}\,\nabla_{\mu}e_{a\,\nu}$, with $\gamma_{cab}=\!-\gamma_{acb}\,$. We will need the following two sets of scalars: the spin coefficients
\begin{fleqn}
\begin{equation}\label{spincoef}
\begin{alignedat}{6}
\kappa=&-\gamma_{311},\quad &\lambda=&\,\gamma_{424}, \quad&\nu=&\,\gamma_{422},\quad &\sigma=&-\gamma_{313},\quad &\alpha=&\,{\textstyle \frac{1}{2}} (\gamma_{124}-\gamma_{344}),\quad &\beta=&\,{\textstyle \frac{1}{2}} (\gamma_{433}-\gamma_{213}),\\
\mu=&\,\gamma_{423},\quad &\rho=&-\gamma_{314},\quad &\pi=&\,\gamma_{421},\quad &\tau=&-\gamma_{312},\quad  &\gamma=&\,{\textstyle \frac{1}{2}}(\gamma_{122}-\gamma_{342}),\quad &\epsilon=&\,{\textstyle \frac{1}{2}} (\gamma_{431}-\gamma_{211}),
\end{alignedat}
\end{equation}
\end{fleqn}
and the Weyl scalars
\begin{equation}\label{NP & Teukolsky_}
\Psi_0= C_{1313}\,,\quad \Psi_1= C_{1213}\,,\quad \Psi_2= C_{1342}\,,\quad \Psi_3= C_{1242}\,,\quad \Psi_4 = C_{2424}\,.
\end{equation}
The complex conjugate of any quantity can be obtain through the replacement $3\leftrightarrow 4$. In a Petrov type D spacetime all Weyl scalars except $\Psi_2$ vanish: $\Psi_0=\Psi_1=\Psi_3=\Psi_4=0\,$. Due to the Goldberg-Sachs theorem this entails $\kappa=\lambda=\nu=\sigma=0\,$. In addition one can set $\epsilon=0$ by choosing $\NPl$ to be tangent to an affinely parametrized null geodesic $\NPl^{\mu} \nabla_{\mu} \NPl_{\nu} =0$.

The various equations of the tetrad formalism can be rewritten using the directional derivatives and the complex scalars of the Newman-Penrose formalism. The Maxwell equations can be treated analogously, one combines the elements of the electromagnetic tensor $F_{\mu\nu}$ into three complex scalars $\phi_0,\,\phi_1,\,\phi_2\,$. Likewise the equations for the components of the Neutrino spinor, $\chi_0$ and $\chi_1$, and the Rarita-Schwinger field, $\Phi_0$ and $\Phi_3$, can be incorporated into the Newman-Penrose formalism.
\subsection{Teukolsky equations}\label{app:NP & Teukolsky_Teukolsky}

The perturbations of spin-$s$ fields in a type D background like the Kerr-AdS geometry are described by the Teukolsky  decoupled  equations, namely by equations (2.12)-(2.15),  (3.5)-(3.8), (B4)-(B5) of \cite{Teukolsky:1973ha}. 
Spin $s=\pm 2, \pm 1, \pm 3/2,\pm 1/2$ describes, respectively, gravitational, electromagnetic, fermionic ($\pm 3/2,\pm 1/2$) perturbations.
These Teukolsky equations for the several spins can be written in a compact form as a pair of equations.
For positive spin field perturbations  the Teukolsky equation is
\begin{align}\label{Teukolskyeqs_pos spin}
\biggl\{&\left[D-\left(2s^+\!-1\right)\epsilon+\bepsilon-2s^+\rho-\brho\right]\left(\Delta+\mu-2s^+\gamma\right)\nonumber\\
&-\left[\delta+\bpi-\balpha-\left(2s^+\!-1\right)\beta-2s^+\tau\right]\left(\bdelta+\pi-2s^+\alpha\right)\nonumber\\
&+{\textstyle \frac{1}{3}}\,s^+\bigl(s^+\!-{\textstyle \frac{1}{2}}\bigr)\left(s^+\!-1\right)\left(2s^+\!-7\right)\Psi_2\biggr\}\,\delta\psi^{(s^+)}=4\pi\mathcal{T}_{(s^+)}\,,\quad s^+=\{1/2,1,3/2,2\},
\end{align}
while  negative spin field perturbations are described by the Teukolsky equation
\begin{align}\label{Teukolskyeqs_neg spin}
\biggl\{&\left[\Delta-\left(2s^-\!+1\right)\gamma-\bgamma-2s^-\mu+\bmu\right]\left(D-2s^-\epsilon-\rho\right)\nonumber\\
&-\left[\bdelta-\btau+\bbeta-\left(2s^-\!+1\right)\alpha-2s^-\pi\right]\left(\delta-\tau-2s^-\beta\right)\nonumber\\
&+{\textstyle \frac{1}{3}}s^-\bigl(s^-\!+{\textstyle \frac{1}{2}}\bigr)\left(s^-\!+1\right)\left(2s^-\!+7\right)\Psi_2\biggr\}\,\delta\psi^{(s^-)}=4\pi\mathcal{T}_{(s^-)}\,,\quad s^-=\{-1/2,-1,-3/2,-2\}.
\end{align}
The explicit form of the source terms $\mathcal{T}_{(s^\pm)}$ is given in~\cite{Teukolsky:1973ha}. 
To make contact with the notation of \eqref{NHgeometry_redef scalars}, note that 
$\delta \psi^{(2)}\equiv \delta \Psi_0,\,\delta \psi^{(-2)}\equiv \delta \Psi_4,\,\delta \psi^{(1)}\equiv \delta\phi_0,\,\delta \psi^{(-1)}\equiv \delta\phi_2,\,\delta \psi^{(\frac{1}{2})}\equiv \delta\chi_0,\,\delta \psi^{(-\frac{1}{2})}\equiv \delta\chi_1,\,\delta \psi^{(\frac{3}{2})}\equiv \delta\Phi_0,\,\delta \psi^{(-\frac{3}{2})}\equiv \delta\Phi_3$. Use of \eqref{NHgeometry_redef scalars} in \eqref{Teukolskyeqs_pos spin} and \eqref{Teukolskyeqs_neg spin} yields \eqref{KerrAdS:TeukEq},  in the Kerr-AdS black hole case, and  \eqref{NHgeometry_master eq}, in the NHEK-AdS geometry case, which are the master equations for the master fields $\Psi^{(s)}$.
 The Teukolsky equations \eqref{Teukolskyeqs_pos spin} and \eqref{Teukolskyeqs_neg spin} are complemented by the Klein-Gordon equation which describes massless scalar perturbations ($s=0$), $\delta\psi^{(0)}\equiv \Psi^{(0)}$,
\begin{equation}\label{NP & Teukolsky_Klein-Gordon eq}
\nabla^2\delta\psi^{(0)}=\frac{1}{\sqrt{-g}}\partial_{\mu}\left(\sqrt{-g}g^{\mu\nu}\partial_{\nu}\delta\psi^{(0)}\right)=0.
\end{equation}

\section{Near-horizon limit of the extremal Kerr-AdS geometry}\label{sec:AppendixNHgeometry}
In this appendix we quickly review the near-horizon limit of the extreme Kerr-AdS black hole \eqref{Kerr:metric}  that generates the NHEK-AdS geometry \eqref{NHgeometry_NH line element}, as first taken in \cite{Lu:2008jk}. We need this explicit limit to discuss the relation between the perturbation frequencies in the full and near-horizon geometries $-$ see discussion associated with \eqref{relationFreq} $-$ and to find the master equation for perturbations in NHEK-AdS in the Poincar\'e frame (see next appendix).  Whether we start from the Kerr-AdS geometry in the rotating Boyer-Lindquist frame or the non-rotating frame will make no difference to the end result.

First we will change to near-horizon coordinates, the associated transformation differs slightly between the two frames. In the rotating frame we make the substitutions
\begin{equation}\label{NH limit_1}
\hat{r}\to r_+\left(1+\lambda\, r'\right)\,, \quad \hat{t}\to \frac{A}{\lambda\, r_+}\,t'\,, \quad \hat{\phi}\to\phi'+\frac{\mathcal{B}}{\lambda\, r_+}\,t'\,,
\end{equation}
while in the non-rotating frame we replace
\begin{equation}\label{NH limit_NH coordinate trafo from non-rotating}
\hat{r}\to r_+\left(1+\lambda\, r'\right)\,, \quad \hat{t}\to \frac{A}{\lambda\, r_+}t'\,, \quad \hat{\varphi}\to\phi'+\frac{B}{\lambda\, r_+}t'\,.
\end{equation}
As the near-horizon geometry is a limit of the extremal Kerr-AdS black hole, the relations~\eqref{full geo_extremality relations} hold. Substitute them into $\Delta_{\hat{r}}$ to find
\begin{equation}\label{NH limit_Delta expansion}
\Delta_{\hat{r}}=V\left(\hat{r}-r_+\right)^2+\mathcal{O}\left(\left(\hat{r}-r_+\right)^3\right)\,, \qquad V=\frac{1+6r_+^2\ell^{-2}-3r_+^4\ell^{-4}}{1-r_+^2\ell^{-2}}\,.
\end{equation}
Only the leading term of $\Delta_{\hat{r}}$ in~\eqref{NH limit_Delta expansion} is relevant for the derivation of the near-horizon geometry. We adjust $A$ such that the metric contains no divergent powers of $\lambda$ and find $A=\frac{\left(r_+^2+a^2\right)}{V}\,$. We now choose $\mathcal{B}$ (or $B$) such that $\phi'$ co-rotates with the horizon and obtain $\mathcal{B}=\frac{a}{V}\left(1-\frac{a^2}{\ell^2}\right)\,$ and $B=\frac{a}{V}\left(1+\frac{r_+^2}{\ell^2}\right)\,$. This difference between $\mathcal{B}$ and $B$ is naturally due to the  coordinate transformation $\hat{\varphi}=\hat{\phi}+\frac{a}{\ell ^2} \hat{t}$ relating the rotating/non-rotating frames of the full geometry.

In a second step we take the near-horizon limit $\lambda\to0$ and find the NHEK-AdS geometry in Poincar\'{e} coordinates
\begin{equation}\label{NHEKpoincare}
ds^2=\frac{\Sigma_+^{\,2}}{V}\left[-r'^2dt'^2+\frac{dr'^2}{r'^2}+\frac{Vd \theta^2}{\Delta_{\theta}}\right]+\frac{\sin^2 \theta \Delta_{\theta}}{\Sigma_+^{\,2}}\left(\frac{2ar_+}{V}\,r'dt'+\frac{\left(r_+^2+a^2\right)}{\Xi}\,d\phi'\right)^2\,,
\end{equation}
where $\Sigma^2_+\!=\!r_+^2+a^2\cos^2\theta$ and $\Delta_{\theta}$ is defined in \eqref{Kerr:metricAUX}. A further coordinate transformation rewrites the NHEK-AdS metric in global coordinates. $AdS_2$ is described by the hyperboloid $Z^2-X^2-Y^2=-1$ in $\Reals^3\,$. Its Poincar\'{e} coordinates $\{r', \,t'\}$ and global coordinates $\{r,\,t\}$ are related via the relations
\begin{equation}\label{NH limit_}
\begin{alignedat}{3}
&X+Z=r'\,, &\qquad& X-Z=\frac{1}{r'}-r't'^2\,, &\qquad& Y=r't'\,,\\
&X=\sqrt{1+r^2}\cos t\,, &\qquad& Y=\sqrt{1+r^2}\sin t\,, &\qquad& Z=r\,.
\end{alignedat}
\end{equation}
From these definitions we find
\begin{equation}\label{NH limit_}
-r'^2dt'^2+\frac{dr'^2}{r'^2}=-\left(1+r^2\right)dt^2+\frac{dr^2}{1+r^2}\,,\qquad r'dt'=rdt+d\gamma\,,
\end{equation}
where
\begin{equation}\label{NH limit_}
\gamma=\ln\Biggl(\frac{1+\sqrt{1+r^2}\sin t}{\cos t+r\sin t}\Biggr)\,.
\end{equation}
To set $g_{r\phi}=0$ we make the final coordinate transformation
\begin{equation}\label{NH limit_}
\theta\rightarrow\theta\,,\qquad \phi'\rightarrow\phi+\frac{2ar_+\Xi\gamma}{(r_+^2 +a^2)V}\,,
\end{equation}
and we find the line element \eqref{NHgeometry_NH line element} of the NHEK-AdS geometry in global coordinates. 
In the limit of a vanishing cosmological constant, which corresponds to $\ell\rightarrow\infty\,$, it reduces to the line element of the NHEK geometry \cite{Bardeen:1999px}.

\section{Master equation for NHEK-AdS in Poincar\'e coordinates\label{masterPoincare}}

The Poincar\'{e} coordinate patch is commonly used in applications of the AdS/CFT correspondence. So, for the sake of completeness, we will present the equivalent of equation~\eqref{NHgeometry_master eq} in Poincar\'{e} coordinates $\{t',r',\theta,\phi'\}$. 

To derive this equation we must apply the near-horizon limit~\eqref{NH limit_NH coordinate trafo from non-rotating} to the master equation  \eqref{KerrAdS:TeukEq} for the Kerr-AdS geometry.
Doing so we find that a spin-$s$ perturbation  $f^{(s)}(t',r',\theta,\phi')$ in the NHEK-AdS background obeys the Teukolsky master equation
\begin{align}\label{PoincarŽ_}
&\frac{V}{r'^2}\,\partial_{t'}^{\,2} f^{(s)}-\frac{4\,a\,\Xi\, r_+}{\left(r_+^2+a^2\right)r'}\,\partial_{t'}\partial_{\varphi'} f^{(s)}+\left(\frac{a^2\left(r_+^2+\ell^2\right)^2\Xi}{\ell^2\left(r_+^2+a^2\right)^2\Delta_{\theta}}-\frac{\Xi^2\,a^2\left(\ell^2-r_+^2\right)}{4 \,r_+^4 \,V}-\frac{\Xi}{\sin^2\theta}\right)\partial_{\varphi'}^{\,2} f^{(s)} \nonumber \\
&-Vr'^{-2s}\,\partial_{r'}\left(r'^{2\left(s+1\right)}\,\partial_{r'}f^{(s)}\right)
-\frac{1}{\sin\theta}\partial_{\theta}\left(\sin\theta\Delta_{\theta}\,\partial_{\theta}f^{(s)}\right)-\frac{2sV}{r'}\,\partial_{t'}f^{(s)} \nonumber \\
&-2\,i\,s\,\Xi\,\cos\theta\left(\frac{1}{\sin^2\theta}+\frac{a^2\left(r_+^2+\ell^2\right)}{\ell^2\left(r_+^2+a^2\right)\Delta_{\theta}}\right)\partial_{\varphi'}f^{(s)}+\biggl[\left(16s^8-120s^6+273s^4\right)\frac{\Sigma_+^{\,2}}{18\ell^2}  \nonumber \\
&+s^2\biggl(\frac{\Xi}{\sin^2\theta}-\frac{\Xi}{\Delta_{\theta}}-\frac{\left(277r_+^2+205a^2\cos^2\theta\right)}{18\ell^2}\biggr)-s\left(1+\frac{a^2}{\ell^2}+\frac{6r_+^2}{\ell^2}\right)\biggr]f^{(s)}=0\,.
\end{align}
To separate the equation we choose the ansatz
\begin{equation*}
f^{(s)}(t',r',\theta,\phi')=F^{(s)}(t',r')S^{(s)}(\theta)e^{im\phi'}
\end{equation*}
and obtain
\begin{equation}
\begin{split}
&\hspace{-1cm}\frac{V}{r'^{\,2}}\,\partial_{t'}^2F^{(s)}-\left(\frac{2sV}{r'}+i\frac{4\,a\,m\,r_+\,\Xi}{\left(a^2+r_+^2\right)r'}\right)\partial_{t'}F^{(s)}\\
&\hspace{1cm}-Vr'^{\,-2s}\,\partial_{r'}\left(r'^{\,2\left(s+1\right)}\,\partial_{r'}F^{(s)}\right)+V\left(\Lambda^{(s)}_{lm}-\frac{7m^2}{4}\right)F^{(s)}=0\,.
\end{split}
\end{equation}
The equation for $S^{(s)}(\theta)$ is identical to the angular equation in global coordinates \eqref{NHgeometry_angular master eq}. The flat limit of our results agrees with the corresponding equations for the NHEK geometry written in Appendix A.2 of \cite{Dias:2009ex}.\footnote{To check this agreement, note that we describe the near-horizon as $\hat{r}\rightarrow r_+\left(1+\lambda r'\right)$, while \cite{Dias:2009ex} uses instead $\hat{r}\rightarrow r_++\lambda r'$.}

\end{appendix}

\bibliographystyle{utphys}
\bibliography{refs}

\providecommand{\href}[2]{#2}\begingroup\raggedright\begin{thebibliography}{10}

\bibitem{Whiting:1988vc}
B.~F. Whiting, ``{Mode stability of the Kerr black hole},''
\href{http://dx.doi.org/10.1063/1.528308}{{\em J.Math.Phys.} {\bfseries 30}
  (1989) 1301}.

\bibitem{Press:1973zz}
W.~H. Press and S.~A. Teukolsky, ``{Perturbations of a Rotating Black Hole. II.
  Dynamical Stability of the Kerr Metric},''
\href{http://dx.doi.org/10.1086/152445}{{\em Astrophys.J.} {\bfseries 185}
  (1973) 649--674}.

\bibitem{Teukolsky:1973ha}
S.~A. Teukolsky, ``{Perturbations of a Rotating Black Hole. 1. Fundamental
  Equations for Gravitational, Electromagnetic and Neutrino Field
  Perturbations},''
\href{http://dx.doi.org/10.1086/152444}{{\em Astrophys. J.} {\bfseries 185}
  (1973) 635--647}.

\bibitem{Bardeen:1999px}
J.~M. Bardeen and G.~T. Horowitz, ``{The Extreme Kerr throat geometry: A Vacuum
  analog of AdS(2) x S**2},''
  \href{http://dx.doi.org/10.1103/PhysRevD.60.104030}{{\em Phys.Rev.}
  {\bfseries D60} (1999) 104030},
\href{http://arxiv.org/abs/hep-th/9905099}{{\ttfamily arXiv:hep-th/9905099
  [hep-th]}}.

\bibitem{Amsel:2009ev}
A.~J. Amsel, G.~T. Horowitz, D.~Marolf, and M.~M. Roberts, ``{No Dynamics in
  the Extremal Kerr Throat},''
  \href{http://dx.doi.org/10.1088/1126-6708/2009/09/044}{{\em JHEP} {\bfseries
  0909} (2009) 044}, \href{http://arxiv.org/abs/0906.2376}{{\ttfamily
  arXiv:0906.2376 [hep-th]}}.
42 pages, 3 figures. v2: references added and minor clarifications.

\bibitem{Dias:2009ex}
O.~J.~C. Dias, H.~S. Reall, and J.~E. Santos, ``{Kerr-CFT and Gravitational
  Perturbations},'' \href{http://dx.doi.org/10.1088/1126-6708/2009/08/101}{{\em
  JHEP} {\bfseries 08} (2009) 101},
\href{http://arxiv.org/abs/0906.2380}{{\ttfamily arXiv:0906.2380 [hep-th]}}.

\bibitem{Marolf:2010nd}
D.~Marolf, ``{The dangers of extremes},''
  \href{http://dx.doi.org/10.1007/s10714-010-1027-z}{{\em Gen.Rel.Grav.}
  {\bfseries 42} (2010) 2337--2343},
\href{http://arxiv.org/abs/1005.2999}{{\ttfamily arXiv:1005.2999 [gr-qc]}}.

\bibitem{Aretakis:2012ei}
S.~Aretakis, ``{Horizon Instability of Extremal Black Holes},''
\href{http://arxiv.org/abs/1206.6598}{{\ttfamily arXiv:1206.6598 [gr-qc]}}.

\bibitem{LuciettiReall2012}
J.~Lucietti and H.~S. Reall, ``{Gravitational instability of an extreme Kerr
  black hole},''
\href{http://arxiv.org/abs/1208.1437}{{\ttfamily arXiv:1208.1437 [gr-qc]}}.

\bibitem{Carter:1968ks}
B.~Carter, ``{Hamilton-Jacobi and Schrodinger Separable Solutions of Einstein's
  Equations},''
{\em Commun. Math. Phys.} {\bfseries 10} (1968) 280.

\bibitem{Dias:2011ss}
O.~J. Dias, G.~T. Horowitz, and J.~E. Santos, ``{Gravitational Turbulent
  Instability of Anti-de Sitter Space},''
  \href{http://arxiv.org/abs/1109.1825}{{\ttfamily arXiv:1109.1825 [hep-th]}}.
5 pages.

\bibitem{Hawking:1999dp}
S.~Hawking and H.~Reall, ``{Charged and rotating AdS black holes and their CFT
  duals},'' \href{http://dx.doi.org/10.1103/PhysRevD.61.024014}{{\em Phys.Rev.}
  {\bfseries D61} (2000) 024014},
\href{http://arxiv.org/abs/hep-th/9908109}{{\ttfamily arXiv:hep-th/9908109
  [hep-th]}}.

\bibitem{Cardoso:2006wa}
V.~Cardoso, O.~J. Dias, and S.~Yoshida, ``{Classical instability of Kerr-AdS
  black holes and the issue of final state},''
  \href{http://dx.doi.org/10.1103/PhysRevD.74.044008}{{\em Phys.Rev.}
  {\bfseries D74} (2006) 044008},
\href{http://arxiv.org/abs/hep-th/0607162}{{\ttfamily arXiv:hep-th/0607162
  [hep-th]}}.

\bibitem{Lu:2008jk}
H.~Lu, J.~Mei, and C.~N. Pope, ``{Kerr/CFT Correspondence in Diverse
  Dimensions},'' \href{http://dx.doi.org/10.1088/1126-6708/2009/04/054}{{\em
  JHEP} {\bfseries 04} (2009) 054},
\href{http://arxiv.org/abs/0811.2225}{{\ttfamily arXiv:0811.2225 [hep-th]}}.

\bibitem{Durkee:2010ea}
M.~Durkee and H.~S. Reall, ``{Perturbations of Near-horizon Geometries and
  Instabilities of Myers-Perry Black Holes},''
  \href{http://dx.doi.org/10.1103/PhysRevD.83.104044}{{\em Phys.Rev.}
  {\bfseries D83} (2011) 104044},
  \href{http://arxiv.org/abs/1012.4805}{{\ttfamily arXiv:1012.4805 [hep-th]}}.

\bibitem{Dias:2010ma}
O.~J. Dias, R.~Monteiro, H.~S. Reall, and J.~E. Santos, ``{A Scalar field
  condensation instability of rotating anti-de Sitter black holes},'' {\em
  JHEP} {\bfseries 1011} (2010) 036,
  \href{http://arxiv.org/abs/1007.3745}{{\ttfamily arXiv:1007.3745 [hep-th]}}.
34 pages/ 13 figures.

\bibitem{Tanahashi:2012si}
N.~Tanahashi and K.~Murata, ``{Instability in near-horizon geometries of
  even-dimensional Myers-Perry black holes},''
\href{http://arxiv.org/abs/1208.0981}{{\ttfamily arXiv:1208.0981 [hep-th]}}.

\bibitem{Chen:2010bh}
B.~Chen and J.~Long, ``{On Holographic description of the Kerr-Newman-AdS-dS
  black holes},'' \href{http://dx.doi.org/10.1007/JHEP08(2010)065}{{\em JHEP}
  {\bfseries 1008} (2010) 065},
\href{http://arxiv.org/abs/1006.0157}{{\ttfamily arXiv:1006.0157 [hep-th]}}.

\bibitem{Guica:2008mu}
M.~Guica, T.~Hartman, W.~Song, and A.~Strominger, ``{The Kerr/CFT
  Correspondence},'' \href{http://dx.doi.org/10.1103/PhysRevD.80.124008}{{\em
  Phys.Rev.} {\bfseries D80} (2009) 124008},
  \href{http://arxiv.org/abs/0809.4266}{{\ttfamily arXiv:0809.4266 [hep-th]}}.

\bibitem{Compere:2012jk}
G.~Comp\`{e}re, ``{The Kerr/CFT correspondence and its extensions: a
  comprehensive review},'' \href{http://arxiv.org/abs/1203.3561}{{\ttfamily
  arXiv:1203.3561 [hep-th]}}.
To be published in Living Reviews in Relativity. Format adapted (76 pages).
  Refs added.

\bibitem{Godazgar:2011sn}
M.~Godazgar, ``{The perturbation theory of higher dimensional spacetimes \'a la
  Teukolsky},'' \href{http://dx.doi.org/10.1088/0264-9381/29/5/055008}{{\em
  Class.Quant.Grav.} {\bfseries 29} (2012) 055008},
\href{http://arxiv.org/abs/1110.5779}{{\ttfamily arXiv:1110.5779 [gr-qc]}}.

\bibitem{Guica:2010ej}
M.~Guica and A.~Strominger, ``{Microscopic Realization of the Kerr/CFT
  Correspondence},'' \href{http://dx.doi.org/10.1007/JHEP02(2011)010}{{\em
  JHEP} {\bfseries 1102} (2011) 010},
\href{http://arxiv.org/abs/1009.5039}{{\ttfamily arXiv:1009.5039 [hep-th]}}.

\bibitem{Bena:2012wc}
I.~Bena, M.~Guica, and W.~Song, ``{Un-twisting the NHEK with spectral flows},''
  \href{http://arxiv.org/abs/1203.4227}{{\ttfamily arXiv:1203.4227 [hep-th]}}.
53 pages, LaTeX.

\bibitem{Gibbons:2004ai}
G.~Gibbons, M.~Perry, and C.~Pope, ``{The First law of thermodynamics for
  Kerr-anti-de Sitter black holes},''
  \href{http://dx.doi.org/10.1088/0264-9381/22/9/002}{{\em Class.Quant.Grav.}
  {\bfseries 22} (2005) 1503--1526},
\href{http://arxiv.org/abs/hep-th/0408217}{{\ttfamily arXiv:hep-th/0408217
  [hep-th]}}.

\bibitem{Caldarelli:1999xj}
M.~M. Caldarelli, G.~Cognola, and D.~Klemm, ``{Thermodynamics of
  Kerr-Newman-AdS black holes and conformal field theories},''
  \href{http://dx.doi.org/10.1088/0264-9381/17/2/310}{{\em Class.Quant.Grav.}
  {\bfseries 17} (2000) 399--420},
\href{http://arxiv.org/abs/hep-th/9908022}{{\ttfamily arXiv:hep-th/9908022
  [hep-th]}}.

\bibitem{Caldarelli:2008ze}
M.~M. Caldarelli, O.~J. Dias, and D.~Klemm, ``{Dyonic AdS black holes from
  magnetohydrodynamics},''
  \href{http://dx.doi.org/10.1088/1126-6708/2009/03/025}{{\em JHEP} {\bfseries
  0903} (2009) 025},
\href{http://arxiv.org/abs/0812.0801}{{\ttfamily arXiv:0812.0801 [hep-th]}}.

\bibitem{Teukolsky:1974yv}
S.~Teukolsky and W.~Press, ``{Perturbations of a rotating black hole. III -
  Interaction of the hole with gravitational and electromagnet ic radiation},''
\href{http://dx.doi.org/10.1086/153180}{{\em Astrophys.J.} {\bfseries 193}
  (1974) 443--461}.

\bibitem{Chambers:1994ap}
C.~M. Chambers and I.~G. Moss, ``{Stability of the Cauchy horizon in Kerr-de
  Sitter space-times},''
  \href{http://dx.doi.org/10.1088/0264-9381/11/4/019}{{\em Class.Quant.Grav.}
  {\bfseries 11} (1994) 1035--1054},
\href{http://arxiv.org/abs/gr-qc/9404015}{{\ttfamily arXiv:gr-qc/9404015
  [gr-qc]}}.

\bibitem{Giammatteo:2005vu}
M.~Giammatteo and I.~G. Moss, ``{Gravitational quasinormal modes for Kerr
  anti-de Sitter black holes},''
  \href{http://dx.doi.org/10.1088/0264-9381/22/9/021}{{\em Class.Quant.Grav.}
  {\bfseries 22} (2005) 1803--1824},
\href{http://arxiv.org/abs/gr-qc/0502046}{{\ttfamily arXiv:gr-qc/0502046
  [gr-qc]}}.

\bibitem{Breuer1977}
R.~A. Breuer, M.~P. Ryan, and S.~Waller, ``{Some Properties of Spin-Weighted
  Spheroidal Harmonics},''
{\em Proc. R. Soc. Lond. A} {\bfseries 22} (1977) 71.

\bibitem{Berti:2005gp}
E.~Berti, V.~Cardoso, and M.~Casals, ``{Eigenvalues and eigenfunctions of
  spin-weighted spheroidal harmonics in four and higher dimensions},''
  \href{http://dx.doi.org/10.1103/PhysRevD.73.024013,
  10.1103/PhysRevD.73.109902}{{\em Phys.Rev.} {\bfseries D73} (2006) 024013},
\href{http://arxiv.org/abs/gr-qc/0511111}{{\ttfamily arXiv:gr-qc/0511111
  [gr-qc]}}.

\bibitem{Suzuki:1999nn}
H.~Suzuki, E.~Takasugi, and H.~Umetsu, ``{Analytic solutions of Teukolsky
  equation in Kerr-de Sitter and Kerr-Newman-de Sitter geometries},''
  \href{http://dx.doi.org/10.1143/PTP.102.253}{{\em Prog.Theor.Phys.}
  {\bfseries 102} (1999) 253--272},
\href{http://arxiv.org/abs/gr-qc/9905040}{{\ttfamily arXiv:gr-qc/9905040
  [gr-qc]}}.

\bibitem{Strominger:1998yg}
A.~Strominger, ``{AdS(2) Quantum Gravity and String Theory},'' {\em JHEP}
  {\bfseries 01} (1999) 007,
\href{http://arxiv.org/abs/hep-th/9809027}{{\ttfamily arXiv:hep-th/9809027}}.

\bibitem{abramowitz1964handbook}
M.~Abramowitz and I.~Stegun, {\em {Handbook of Mathematical Functions with
  Formulas, Graphs, and Mathematical Tables}}.
\newblock Applied Mathematics Series. U.S. Govt. Print. Off., 1964.

\bibitem{Friedman:1978}
J.~Friedman, ``{Ergosphere instability},'' {\em Commun. Math. Phys.} {\bfseries
  63} (1978) 243.

\bibitem{friedman}
J.~Friedman, ``{Ergosphere instability},''
  \href{http://arxiv.org/abs/1203.3561}{{\ttfamily arXiv:1203.3561 [hep-th]}}.
To be published in Living Reviews in Relativity. Format adapted (76 pages).
  Refs added.

\bibitem{Cohen19755}
J.~M. Cohen and L.~S. Kegeles, ``{Space-time Perturbations},''
  \href{http://dx.doi.org/DOI: 10.1016/0375-9601(75)90583-6}{{\em Physics
  Letters A} {\bfseries 54} no.~1, (1975) 5--7}.

\bibitem{Kegeles:1979an}
L.~S. Kegeles and J.~M. Cohen, ``{Constructive Procedure for Perturbations of
  Space-times},''
\href{http://dx.doi.org/10.1103/PhysRevD.19.1641}{{\em Phys. Rev.} {\bfseries
  D19} (1979) 1641--1664}.

\bibitem{Chrzanowski:1975wv}
P.~L. Chrzanowski, ``{Vector Potential and Metric Perturbations of a Rotating
  Black Hole},''
\href{http://dx.doi.org/10.1103/PhysRevD.11.2042}{{\em Phys. Rev.} {\bfseries
  D11} (1975) 2042--2062}.

\bibitem{Stewart:1978tm}
J.~M. Stewart, ``{Hertz-Bromowich-Debye-Whittaker-Penrose Potentials in General
  Relativity},''
\href{http://dx.doi.org/10.1098/rspa.1979.0101}{{\em Proc. Roy. Soc. Lond.}
  {\bfseries A367} (1979) 527--538}.

\bibitem{Wald:1978vm}
R.~M. Wald, ``{Construction of Solutions of Gravitational, Electromagnetic, or
  Other Perturbation Equations from Solutions of Decoupled Equations},''
\href{http://dx.doi.org/10.1103/PhysRevLett.41.203}{{\em Phys. Rev. Lett.}
  {\bfseries 41} (1978) 203--206}.

\bibitem{Price:2006ke}
L.~R. Price, K.~Shankar, and B.~F. Whiting, ``{On the Existence of Radiation
  Gauges in Petrov Type II Spacetimes},''
  \href{http://dx.doi.org/10.1088/0264-9381/24/9/014}{{\em Class.Quant.Grav.}
  {\bfseries 24} (2007) 2367--2388},
  \href{http://arxiv.org/abs/gr-qc/0611070}{{\ttfamily arXiv:gr-qc/0611070
  [gr-qc]}}.

\end{thebibliography}\endgroup
\end{document}